\documentclass[a4paper,11pt]{article}

\usepackage[a4paper,left=2.73cm,right=2.7cm,top=3cm,bottom=3.5cm]{geometry}
\usepackage{amsmath,amssymb,graphicx,hyperref}
\usepackage[all]{xy}

\numberwithin{equation}{section}

\begin{document}

\begin{titlepage}

\hfill{ICCUB-13-247}

\vspace{1cm}
\begin{center}
{\huge{\bf Transport properties of spacetime-filling branes}}

\vskip 45pt
{\Large \bf Javier Tarr\'\i o}

\vskip 10pt
{Departament de F\'\i sica Fonamental and Institut de Ci\`encies del Cosmos, Universitat de Barcelona, Mart\'\i\  i Franqu\`es 1, ES-08028, Barcelona, Spain.}\\

\vskip 10pt
{ j.tarrio@ub.edu}
\end{center}

\vspace{10pt}
\abstract{\normalsize
A model  consisting of (d+1)-dimensional gravity coupled to  spacetime filling charged branes is used to study the effects of backreaction. 
The charged black holes arising from this simple model reflect the non-linearity of the gauge field and are thermodynamically stable.
By analysing fluctuations of the system  we corroborate that at low values of the temperature (or large chemical potential) backreaction effects from the branes  are dominant.
We also provide a generalisation of the Iqbal and Liu strategy to calculate the DC conductivity, in which a mass term for the gauge field fluctuation is included.
This mass term gives the value of the residue of the pole at zero frequency in the imaginary part of the AC conductivity, as well as the running of the DC conductivity with the bulk radius.
}

\end{titlepage}

\tableofcontents

\hrulefill
\vspace{15pt}

\section{Introduction}

The gauge/gravity correspondence \cite{Maldacena:1997re} is a widely used tool which allows  to understand certain strongly coupled theories.
Top-down theories arising from string theory have been studied extensively to give ballpark estimates of strongly coupled phases of QCD \cite{CasalderreySolana:2011us} and condensed matter systems \cite{Benini:2012iq}.

The characteristics of a top-down approach are the knowledge of the field theory under study and the spectrum of operators (referred to as the holographic dictionary). 
Finding consistent solutions  describing lower dimensional field theories with given required properties is not a simple task. Many times, the holographic dual used to provide estimates to a field theory does not describe all the expected properties of the physical system under study.
In these cases the holographic model being used should be replaced by a different top-down model that succeeds to incorporate the properties under study. 
An alternative approach is to build up a model that includes those properties by construction, fitting the different parameters by comparing to the experiments.
This bottom-up viewpoint has the inconvenience that the number and form of operators and states of the theory is unknown: a field theory that is stable under the deformations of its Lagrangian by a set of operators can be unstable under the deformation by a different set.

One advantage of bottom-up models is that they are a useful tool to determine what are the important characteristics that have to be present in a top-down approach to understand features of the physical properties under study. 
There are several examples where a deep knowledge of bottom-up models helped to identify consistent truncations of string theory where properties of interest were to be found.
To name a few, holographic superconductivity was first constructed using a minimal set of matter fields and gravity \cite{Hartnoll:2008vx}, and only later string theory embeddings were studied \cite{Denef:2009tp,Gubser:2009qm,Gauntlett:2009dn}. The same happened for theories with Lifshitz or Schr\"odinger scaling, proposed in \cite{Kachru:2008yh} and \cite{Son:2008ye} respectively and then obtained from string theory in \cite{Herzog:2008wg,Maldacena:2008wh,Adams:2008wt}.
A different application of bottom-up models is a phenomenological approach.
In these cases the model is studied and improved without worrying much about string theory embeddings, but about the physical consequences extracted from it. 
An example is the study of theories with Lifshitz scaling and hyperscaling violation from a theory of Einstein gravity coupled to a dilatonic scalar and a U(1) gauge field \cite{Taylor:2008tg,Charmousis:2010zz}, which allows to study theories with broken Lorentz symmetry under a better control than the top-down models, or the `Improved Holographic QCD' program \cite{Gursoy:2010fj,Jarvinen:2011qe,Alho:2012mh} where a holographic dual with the properties of QCD is being constructed.

\vspace{15pt}

The work presented in this paper falls in the latter category. The model under consideration, although based on the type of Lagrangians  obtained in top-down approaches,  consists of a simplified version  describing the interaction between adjoint and fundamental matter
\begin{equation}\label{eq.model}
S = S_{adj} + S_{fun}  \ ,
\end{equation}
where the fundamental matter is charged under a U(1) vector.
Understanding the implications of charged fundamental matter onto the adjoint sector is the main motivation of this work.
The adjoint matter is described in the saddle-point approximation by a supergravity, consisting of a graviton, a dilaton and  p-forms. A dimensional reduction usually implies the existence of several matter fields, and a non trivial potential for the lower-dimensional scalars. A simple choice for the adjoint sector is to consider the matter sector to consist of just one scalar
\begin{equation}\label{eq.modeladj}
S_{adj} = \frac{1}{2\kappa^2} \int d^{d+1} x \sqrt{-g} \left( R - \frac{1}{2} \partial_\mu\phi \partial^\mu\phi - V(\phi) \right)  \ .
\end{equation}
$R$ is the Ricci scalar associated to the metric $g$, and $\tfrac{1}{2\kappa^2}$ gives information about the number of degrees of freedom of the adjoint sector.
As an example, when $d=4$ and $V=-\tfrac{12}{L^2}$, this action describes the dimensional reduction of type IIB supergravity on a 5-sphere of radius $L$.
There is a solution consisting of an $AdS_5$ spacetime of radius $L$, with trivial dilaton, that describes holographically ${\cal N}=4$ SYM theory \cite{Maldacena:1997re}.
In this case $\tfrac{L^3}{2\kappa^2} \sim N^2$.
Another example  is that of ABJM theory when $d=3$ \cite{Aharony:2008ug}. There is an $AdS_4$ solution of radius $L$ and trivial dilaton for $V=-\tfrac{6}{L^2}$, originating from a reduction of M-theory on the orbifold $\mathbb{S}^7/\mathbb{Z}_k$. 
Then  $\tfrac{L^2}{2\kappa^2} \sim \sqrt{k}\, N^{3/2}$.

Fundamental matter is modelled from string theory by the presence of open strings with one end on the set of branes describing the adjoint matter and one end on a second stack of branes, whose action is given by a Dirac-Born-Infeld (DBI) and a Wess-Zumino (WZ) contribution. 
In dimensional reductions the latter contribution usually appears in a topological sector involving differential forms, or is taken care of in the reduction ansatz. 
Ignoring any contribution from higher order forms, as has been done in the adjoint sector \eqref{eq.modeladj}, the part of the action describing the fundamental sector is given by a DBI-like term corresponding to a stack of coincident, spacetime-filling branes
\begin{equation}\label{eq.modelfun}
S_{fun} = - \frac{T_b}{2\kappa^2} \int d^{d+1} x  \, Z_1(\phi) \sqrt{-\det\left( Z_2(\phi) \, g+ \lambda F \right)}  \ .
\end{equation}
In this action $\tfrac{T_b}{2\kappa^2}$ is the tension of the brane, although in the rest of the paper this name is used for the dimensionless combination $T_b \, L^2$, with $L$ the curvature characterising the spacetime.
The tension of the brane gives information about the ratio between the degrees of freedom in the fundamental and in the adjoint representations. 
In the case of ${\cal N}=4$ SYM, $\tfrac{T_b}{2\kappa^2} L^5 \sim \lambda_{'t}\tfrac{N_{fun}}{N}$, with $\lambda_{'t}$ the 't Hooft coupling. 
$Z_1$ and $Z_2$ are generic dilatonic couplings to the brane.
Finally, $\lambda$ is the string tension. 
At this stage it can be reabsorbed in the diagonal $U(1)\subset U(N_{fun})$ gauge field whose field strength is $F=dA$, due to the absence of charged matter in the system.
In this paper it is kept as a bookkeeping device.

\vspace{15pt}

The tension of the brane controls the backreaction of the fundamental matter onto the adjoint sector.
In the limit $T_b \, L^{2}\to 0$, the adjoint and fundamental terms in \eqref{eq.model} decouple, and the probe approximation of \cite{Karch:2002sh} is recovered. 
However, in \cite{Hartnoll:2009ns} it was argued that in certain cases, for non trivial values of the U(1) field in \eqref{eq.modelfun}, this approximation is not valid: the backreaction of the fundamental matter is generically important in the IR description of the theory.
This was confirmed in the specific case of the intersection of a stack of D3-branes with smeared D7-branes at finite charge density \cite{Bigazzi:2013jqa}, where a perturbative solution in $T_b\, L^{2}$ was constructed and shown to break down in the interior of the bulk geometry.\footnote{Notice that in \cite{Bigazzi:2013jqa} the action \eqref{eq.model} is a good approximation at leading order in $T_b L^{2}$, but at higher orders several other matter fields were needed.} 
Whether the probe approximation is valid or not depends strongly on the kind of observables under consideration. In the absence of worldvolume gauge fields the perturbative solution to \eqref{eq.model} is enough to calculate the entanglement entropy of probe branes \cite{Chang:2013mca,Jensen:2013lxa}.

Assuming isotropy the gauge field can be partially gauge-fixed to  $A=A_t(r) dt$. 
This field enters only via derivatives in the action \eqref{eq.modelfun}, and is solved  in terms of a constant of integration denoted by the dimensionless parameter $\rho$
\begin{equation}\label{eq.Atsolution}
\lambda\, A_t' = \rho \frac{Z_2 \sqrt{|g_{tt}| g_{rr}}}{\sqrt{\rho^2 + Z_1^2 Z_2^{d-1}g_{xx}^{d-1} }} \ .
\end{equation}
This solution is plugged into the equations of motion arising from \eqref{eq.model} to eliminate $A_t'$. 
Equivalently, it is possible to perform a Legendre transform in the action to eliminate the gauge field in terms of $\rho$, and obtain the equations of motion from\begin{equation}\label{eq.system}
S = \frac{1}{2\kappa^2} \int d^{d+1}x \sqrt{-g} \left(  R - \frac{1}{2} \partial_\mu \phi \partial^\mu \phi - V(\phi) -  T_b \, Z_2(\phi) \sqrt{ Z_1(\phi)^2 Z_2(\phi)^{d-1} + g_{xx}^{1-d} \rho^2} \right) \ .
\end{equation}
This action is not explicitly covariant due to the appearance of the $g_{xx}^{1-d}$ factor inside the square root.
The Legendre transform is equivalent to a change of boundary condition in the variational problem.
Action \eqref{eq.model}, with the addition of the Gibbons-Hawking term, is appropriate for a configuration in which the  fields at the boundary, in particular the gauge field $A_t(r_\infty)$, are kept fixed. 
On the other hand, \eqref{eq.system} is appropriate when $\rho$ is the quantity kept fixed.
This will be important later, when the thermodynamics of the different solutions is studied.

The radicand in the square root of \eqref{eq.system} provides a measure of large or small amplitude of the field strength \eqref{eq.Atsolution}. 
A small value of $F$ is defined when the $\rho$-dependent part of the radicand is negligible
\begin{equation}\label{eq.smallrho}
Z_1^2 \, Z_2^{d-1} g_{xx}^{d-1} \gg    \rho^2  \ .
\end{equation}
This is in principle a radial-dependent statement. 
Provided it holds for every value of $r$, the  action \eqref{eq.system} is approximated by
\begin{equation}\label{eq.systemsmallrho}
S_{\text{small }\rho} =  \frac{1}{2\kappa^2} \int d^{d+1}x \sqrt{-g} \left(  R - \frac{1}{2} \partial_\mu \phi \partial^\mu \phi - \tilde V(\phi) -  \frac{T_b}{2} \tilde Z \, g_{xx}^{1-d}\, \rho^2  \right) \ ,
\end{equation}
with potential
$
 \tilde V(\phi) \equiv V + T_b \, Z_1 Z_2^{\tfrac{d+1}{2}} 
$
and dilatonic coupling
$
\tilde Z(\phi) = Z_1^{-1}Z_2^{\tfrac{3-d}{2}} 
 $.
The effective action \eqref{eq.systemsmallrho} is  the Legendre-tranformed version of an Einstein-Maxwell-Dilaton (EMD) system, where only the time-component of the gauge potential is turned on, and expressed it in terms of the conjugate variable $\rho$. 
This  type of systems were studied in detail in \cite{Goldstein:2009cv,Charmousis:2010zz,Gouteraux:2011ce}. 
For a non trivial potential there exist solutions with hyperscaling-violating metric and Lifshitz scaling, as well as a log-behaved dilaton. 
When the potential is constant there is a solution of the Lifshitz kind, but with no hyperscaling violation \cite{Taylor:2008tg}.

On the other hand, if $\rho$ is large 
\begin{equation}\label{eq.largerho}
Z_1^2 \, Z_2^{d-1} g_{xx}^{d-1} \ll    \rho^2  \ .
\end{equation}
the effective action is
\begin{equation}\label{eq.systemlargerho}
S_{\text{large }\rho} = \frac{1}{2\kappa^2} \int d^{d+1}x  \sqrt{-g} \left(  R - \frac{1}{2} \partial_\mu \phi \partial^\mu \phi - V(\phi) -  \tilde T_b \,  Z_2(\phi) \, g_{xx}^\frac{1-d}{2}  \right) \ ,
\end{equation}
where  $\tilde T_b=T_b\, \rho$. 
The last term in \eqref{eq.systemlargerho} can be rewritten as $-\tfrac{\tilde T_b}{2\kappa^2}\int Z_2 \sqrt{-g_{tt} \, g_{rr}}$, which is a Nambu-Goto (NG) type of action, with strings stretched along the radial direction and arbitrary dilatonic coupling.

\vspace{15pt}

In sections \ref{sec.UV} and \ref{sec.IR}, different solutions to the equations of motion resulting from \eqref{eq.model}, first presented in \cite{Pal:2012zn}, are reviewed, although some  conclusions differ from those drawn in that paper.
In particular, the analysis presented here shows explicitly the thermodynamic stability of the solution.

Section \ref{sec.UV} considers the case with $Z_1=Z_2=1$, and a bare, negative, constant potential $V=2\tilde \Lambda = - \tfrac{d(d-1)}{\tilde L^2}$.
The dilaton is trivial and the metric consists of an asymptotically AdS,  charged, black hole solution.
This solution reduces to the AdS-Reissner-Nordstr\"om one in the low $\rho$ regime, and in many ways it can be considered as a non-linear counterpart.

The potential $V$ is \emph{not} related to the AdS radius, $L$, in  the usual way from gravity with a cosmological constant, $V\sim L^{-2}$.
To see this consider the neutral case, $\rho=0$. The DBI term in the action contributes to the cosmological constant  $2\Lambda=2\tilde \Lambda+T_b$.
Imposing that both the bare cosmological constant, $\tilde \Lambda$, and the dressed one, $\Lambda$, are negative,  this shift determines a maximum allowed value for the tension of the brane, $T_b<2|\tilde \Lambda|$.
This condition restricts the maximum allowed value for the tension or the brane in top-down approaches where the action \eqref{eq.model} is a valid truncation.
For the bottom-up case considered in this paper the condition $T_b<2|\tilde \Lambda|$ will always be satisfied by construction of the solutions.

At fixed value of $T_b$ there is a maximum value of the parameter $\rho=\rho_{ext}$, for which the temperature vanishes.
Since $\rho$ is proportional to the charge, this is a signal of the competition between gravitational attraction and electric repulsion, which is exactly balanced when the temperature of the black hole vanishes.
As in the AdS-Reissner-Nordstr\"om case, at zero temperature the solution has a positive value of the entropy density,  indicating that the solution does not describe the true ground state of the theory.
Extremal black holes where the large $\rho$ approximation \eqref{eq.largerho} is valid exist for small values of the tension of the brane $T_b\ll\sqrt{2} |\tilde \Lambda|$. 
On the contrary, for $T_b\gg\sqrt{2} |\tilde \Lambda|$ the small $\rho$ approximation \eqref{eq.smallrho} is accurate.

Section \ref{sec.IR} deals  with non-trivial values of the  functions $Z_1$, $Z_2$ and $V$, which  are chosen to be simple exponentials  
\begin{equation}
Z_1 = e^{-\alpha\, \phi} \ , \qquad Z_2 = e^{\beta\, \phi} \ , \qquad V = 2\tilde \Lambda\,  e^{m\, \phi} \ .
\end{equation}
These simple functions allow to find scaling solutions\begin{equation}\label{eq.hvlmetric}
ds^2  = r^{-2\gamma} \left( - r^{2z} f(r) dt^2 + r^2 d\vec x^2 + \frac{dr^2}{r^2 f(r)} \right) \ ,
\end{equation}
where $z$ is the dynamic exponent that characterises the breaking of Lorentz symmetry, and $\gamma$ is an overall scale factor that determines the hyperscaling violation, sometimes found in the literature as $\gamma=\theta/(d-1)$.
These two exponents are determined from the parameters in the action \eqref{eq.system}.
The function $f$ is a blackening factor for a neutral black hole, despite the existence of a non-vanishing gauge field \eqref{eq.Atsolution} in the solution.
The hyperscaling violating metrics for this system exist only in the canonical ensemble, where the value of $\rho$ is kept fixed, and  are supported by the existence of the gauge and dilaton fields. 
Any value of $\rho$ is allowed, but only if specific relations between the parameters in the action ($\alpha$, $\beta$, $m$, $\tilde \Lambda$, $T_b$ and $\rho$) hold.

\begin{figure}[tb]
\begin{center}
\includegraphics[scale=0.6]{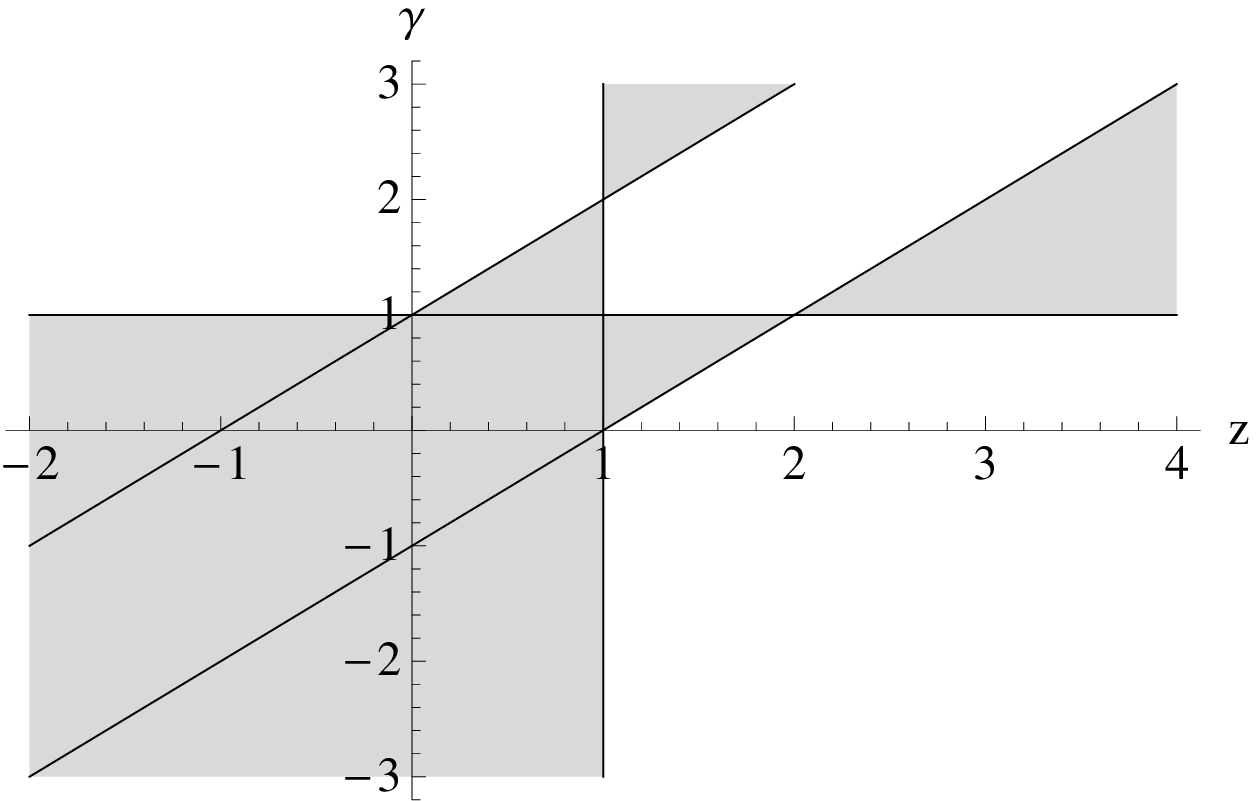}
\includegraphics[scale=0.6]{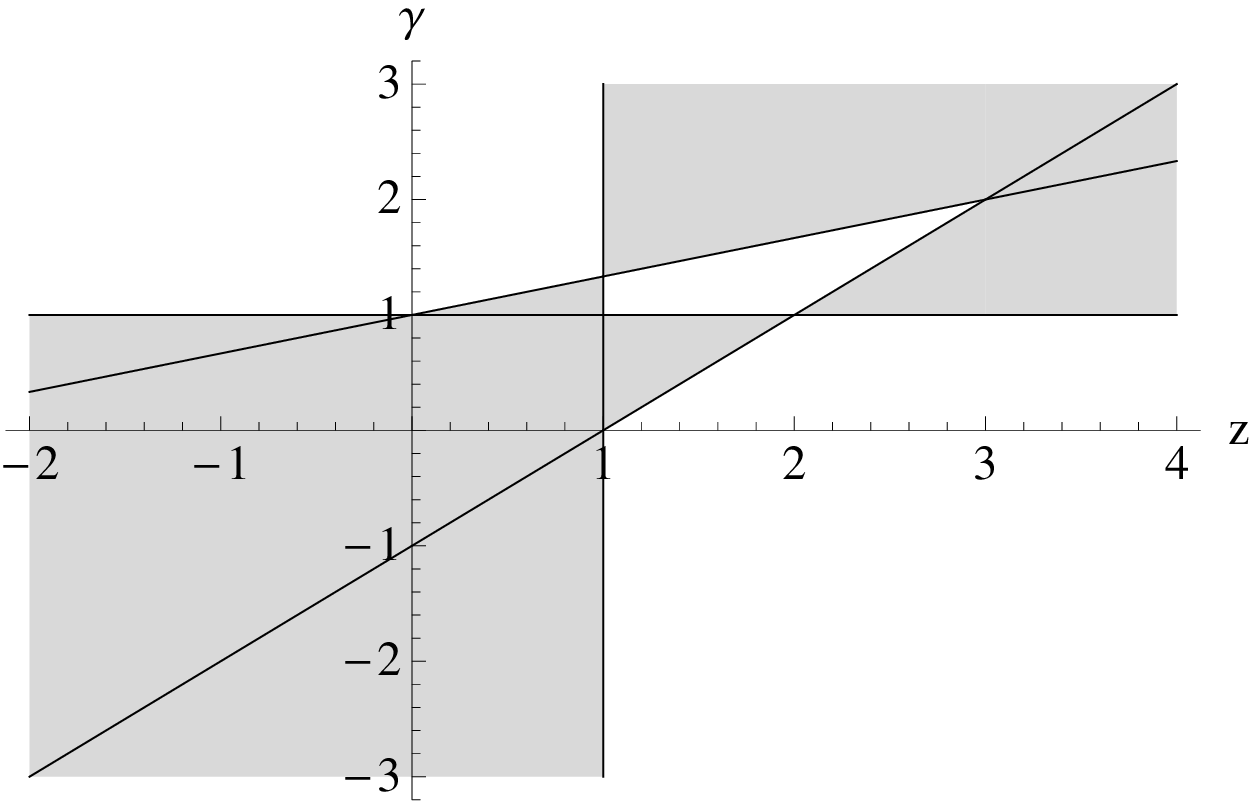}
\caption{Regions where the NEC is  satisfied (light regions) in the $(z,\gamma)$ plane for $d=2$ (left) and $d=4$ (right). The limiting lines are $z=1$, $\gamma=1$, $\gamma=z-1$ and $\gamma=\tfrac{z+d-1}{d-1}$. Only the bottom-right sector, where $z-\gamma\geq1$, is physical in the solutions studied here.}\label{fig.NEC}
\end{center}
\end{figure}

The null energy condition (NEC) $T_{\mu\nu}u^\mu u^\nu\geq0$ with $u^\mu$ a null vector, leads to the inequalities
\begin{equation}\label{eq.NEC}
(z-1)(z+(d-1)(1-\gamma))\geq 0 \ , \quad (\gamma-1)(1+\gamma-z) \geq 0 \ .
\end{equation}
There are three different regions in the $(z,\gamma)$ plane,  shown in figure \ref{fig.NEC}, satisfying these inequalities.
As  shown later, only  the region with $\gamma<1$ and $z\geq1$ is physical from thermodynamic considerations.

Section \ref{sec.fluctuations} contains  the study of the fluctuations of the system \eqref{eq.model}, from where different transport coefficients of the field theory are extracted.
Three independent fluctuation channels are identified according to their transformation under the conserved little group.
The study of the retarded current-current correlator gives information about the conductivity of the system.
At zero frequencies there is a delta function in the real part of the conductivity, signalling momentum conservation.
The behaviour of the conductivity with the frequency is also studied, identifying Drude-like peaks at low frequencies in some regions of parameter space, namely the tension of the brane and ratio between temperature and chemical potential of the system.

The study of quasinormal modes (QNM) for the UV-complete solution of section \ref{sec.UV} is started, focusing on hydrodynamic modes: those whose energy vanishes as a power of the momentum, $\omega \sim q^n$.
In particular there are two types of hydrodynamic QNM, according to their behaviour at low momenta. The diffusive type is defined via the dispersion relation
\begin{equation}
\omega = - i \, {\cal D}\, q^2 + {\cal O}(q^3) \ ,
\end{equation}
and the sound type by the dispersion relation
\begin{equation}
\omega = c\, q - i\, \Gamma \, q^2 + {\cal O}(q^3) \ .
\end{equation}
For the range of parameters investigated in this paper, the numerical dispersion relation of the sound and diffusive modes are always well described by the hydrodynamic result in which $c=c_s$ is the speed of sound, $\Gamma=\tfrac{d-2}{d-1} \tfrac{\eta}{{\cal E} + P}$ the attenuation factor, and the diffusion constant is given by the susceptibility matrix and the DC conductivity (see equation \eqref{eq.ERel}). In these expressions $\cal E$ the internal energy of the system, $P$ the pressure and $\eta$ the shear viscosity.

In section \ref{sec.conclusions}  conclusions and outlook for extensions of the work presented here are provided. 
There is also appendix \ref{app.equations}, where the  set of scalar fluctuations of the system \eqref{eq.model} is written explicitly.

\section{Charged UV solution} \label{sec.UV}

In this section a solution the equations of motion to the action \eqref{eq.model} in the case in which 
\begin{equation}
Z_1=1 \ , \quad Z_2=1\ , \quad V=2\tilde \Lambda=-\tfrac{d(d-1)}{\tilde L^2} \ ,
\end{equation}
is provided.
This choice for the potential and dilatonic couplings implies that the dilaton is a constant.

The metric reads \cite{Pal:2012zn}
\begin{equation}\label{eq.UVsolution}
ds^2  = - \frac{r^2}{L^2} f(r) dt^2 + \frac{r^2}{L^2} d \vec x^2_{d-1} + \frac{L^2}{r^2} \frac{dr^2}{f(r)} \ ,
\end{equation}
with $L$ the radius of the asymptotic (near the boundary $r\to\infty$) AdS space, and the blackening function 
\begin{equation}\label{eq.blackening}
f(r)  = 1 +\frac{T_b\, L^2}{d(d-1)} - \frac{T_b \, L^2}{d-1} \, |\rho|\, \frac{L^{d-1}}{r^{d-1}} \,{}_2F_1\left( - \frac{1}{2} , \frac{1}{2d-2}  ; \frac{2d-1}{2d-2} ; - \frac{1}{\rho^2} \frac{r^{2(d-1)}}{L^{2(d-1)}} \right)  - \frac{k_d}{r^d} \ ,
\end{equation}
where $k_d$  a constant of integration. 
This constant of integration is related to the largest root of the blackening function, $r_+$, by solving $f(r_+)=0$, and notice that  at large radius $f\to 1$.
The solution is UV-complete, meaning that the metric is asymptotically AdS, and the isometries of a CFT are recovered in the ultraviolet.

The relation between the bare radius, $\tilde L$, appearing in the action and the dressed one, $L$ appearing in the solution \eqref{eq.UVsolution} is
\begin{equation}\label{eq.baredressed}
L^2 = \tilde L^2 \frac{d (d-1)}{d(d-1) - T_b\, \tilde L^2} \ ,
\end{equation}
where the condition $T_b<2|\tilde \Lambda|$ is clearly equivalent to $L^2>0$.
Notice that the blackening function \eqref{eq.blackening} is given in terms of the dimensionless combination $T_b\, L^2\in [0,\infty)$, and that we will use it in the analysis below.
It is always possible to translate back to $T_b \, \tilde L^2$ with equation \eqref{eq.baredressed}, and this is done in some specific cases below.

For simplicity, it is also convenient  to work with the quantity
\begin{equation}\label{eq.varrho}
\varrho= \rho \, \frac{L^{d-1}}{r_+^{d-1}} \ .
\end{equation}
 
 \vspace{15pt}

The temperature and entropy are obtained in a straightforward way: imposing the absence of conical singularities of the euclideanized geometry gives the temperature, whereas the area of the black hole event horizon gives the Bekenstein-Hawking entropy
\begin{equation}\label{eq.temperature}
T= \frac{d\, r_+}{4\pi L^2} \left[ 1 + T_b\,L^2   \frac{\left( 1 - \sqrt{ 1+\varrho^2 } \right)}{ d (d-1)} \right] \ , \qquad S = \frac{2\pi}{\kappa^2} V_{d-1} \frac{r_+^{d-1}}{L^{d-1}} \ ,
\end{equation}
with $V_{d-1}$ the volume of the $d-1$ dimensional spatial part of the manifold. 
In $\mathbb{R}^{d-1}$ this volume is infinity, but intrinsic quantities like $s=S/V_{d-1}$ are finite.

Substituting the solution in \eqref{eq.Atsolution} and imposing the boundary condition $A_t(r_+)=0$ the profile of the gauge field is obtained. The asymptotic behaviour near the boundary reads
\begin{equation}
 A_t \approx    A_t(r_\infty) - \frac{\varrho\, r_+^{d-1}}{\lambda (d-2)r^{d-2}} + {\cal O} \left( r^{4-3d} \right)   \quad \text{for }d>2 \ .
\end{equation}
The asymptotic value of the gauge field is identified holographically with the chemical potential
\begin{equation}\label{eq.mu}
\mu = A_t(r_\infty) = \frac{\varrho}{\lambda} \frac{r_+}{(d-2)} \, {}_2F_1 \left( \frac{1}{2}, \frac{d-2}{2d-2} ; \frac{3d-4}{2d-2} ; - \varrho^2   \right)  \qquad \text{for }d>2 \ .
\end{equation}
The value of the charge associated to this chemical potential is
\begin{equation}\label{eq.charge}
Q \equiv  \int d^{d-1}\vec{x}\frac{\delta S}{\delta A_t'}  = \frac{T_b\, \lambda}{2\kappa^2} V_{d-1}\, \rho\ .
\end{equation}

\vspace{15pt}

The charge susceptibility follows from the definitions just given 
\begin{equation}\label{eq.suscepdef}
\chi = \left( \frac{\partial Q}{\partial \mu} \right)_T \ ,
\end{equation}
where the variation with respect to the chemical potential must be taken at fixed temperature.
The solution is given parametrically  in terms of  $r_+$ and $\varrho$, instead of the temperature and the chemical potential.
Therefore, a variation of the temperature is written as
$
\delta T = \tfrac{\partial T}{\partial r_+} \delta r_+ + \tfrac{\partial T}{\partial \varrho} \delta \varrho \ .
$
Imposing $\delta T=0$ gives  a relation between $\delta r_+$ and $\delta \varrho$, which plugged into \eqref{eq.suscepdef} gives
\begin{equation}\label{eq.suscep}
\chi =  \frac{S}{4\pi} \lambda^2 \, T_b \, L^2\left( \frac{ 4\pi \,T }{d(d-1)\sqrt{1+\varrho^2} + T_b \, L^2 \left( \sqrt{1+\varrho^2} - 1 + (d-2) \varrho^2 \right)} + \frac{\lambda}{L^2 (d-1)} \frac{\mu}{\varrho}  \right)^{-1} \ . 
\end{equation}
The final result has been written showing explicitly that it is non-negative.
 In particular this result is non-singular, meaning that the Einstein-Dilaton-DBI system describes compressible states of matter.
Notice that the temperature and chemical potential dependence is obscured by the fact that the entropy and $\varrho$ are themselves functions of  $\mu$ and  $T$.

\subsection{Thermodynamics in the grand-canonical ensemble}

The free energy associated to the solution in the grand-canonical ensemble is related to the on-shell, euclidean action \eqref{eq.model} via the relation $\Omega = - T \, I_{E}$.
The internal energy is obtained from the time-time component of the Brown-York energy-momentum tensor, varying the on-shell action \eqref{eq.model} with respect to the boundary metric.
These calculations diverge, and the standard holographic procedure to regularise the results consists of the addition of a set of counterterms to the action, an evaluation at a cutoff that cancels the divergent behaviour, and a posterior limit in which  the  cutoff  is sent to the boundary.

In the grand-canonical ensemble the action \eqref{eq.model} is supplemented with two pieces
\begin{equation}
S_{adj} + S_{fun} \to S_{adj} + S_{fun} + S_{GH} + S_{ct} \ ,
\end{equation}
where $S_{GH}$ is the Gibbons-Hawking term included to define a correct variational problem in the gravity sector
\begin{equation}
S_{GH} = \frac{1}{\kappa^2} \int_{r_\Lambda} d^d x \, \sqrt{-\gamma} \, K \ ,
\end{equation}
with $K$ the trace of the extrinsic curvature $K_{\mu\nu}=\nabla_{(\mu} n_{\nu)}$, $n_\mu$ a vector normal to the constant $r$ surfaces with unit norm, and $\gamma_{ab}$ the induced metric at the cutoff surface $r=r_\Lambda$.

 $S_{ct}$ is a volume counterterm \cite{Balasubramanian:1999re}
\begin{equation}
S_{ct} = - \frac{1}{\kappa^2} \int_{r_\Lambda} d^d x \,\frac{d-1}{L} \sqrt{-\gamma} \ .
\end{equation}
These are the only counterterms needed to regularise the on-shell action and stress-energy tensor 
\begin{equation}
T_{ab} = - \frac{1}{\kappa^2} \frac{\sqrt{\gamma_{xx}^{d-1}}}{\sqrt{-\gamma_{tt}}} \left( K_{ab} - K\,\gamma_{ab} - \frac{d-1}{L} \gamma_{ab} \right) \ ,
\end{equation}
from where the energy of the field theory, ${\cal E}=T_{tt}$, and pressure, $P=T_{xx}$, follow.

The necessary components of the extrinsic curvature are
\begin{equation}
K_{tt} = - \frac{r^2 \sqrt{f} \left( 2f+rf' \right)}{2L^3} \ , \quad K_{xx} = \frac{r^2 \sqrt{g}}{L^3} \ , \quad K = \frac{2\,d\,f + rf'}{2L\sqrt{f}} \ ,
\end{equation}
from where the following results follow
\begin{equation}\label{eq.ADMUV}
{\cal E}  =  \frac{1}{2\kappa^2} V_{d-1} \frac{(d-1)}{L^{d+1}} m \ , 
\end{equation}
and
\begin{equation}
P  = \frac{\cal E}{d-1} \ ,
\end{equation}
where the new parameter $m$ is defined as 
\begin{equation}\label{eq.mcoeff}
m= k_d-\frac{T_b\, L^2}{2(d-1)\sqrt{\pi}} L^{d}\, |\rho|^\frac{d}{d-1} \Gamma\left( - \frac{d}{2d-2} \right)  \Gamma\left( \frac{2d-1}{2d-2} \right) \geq 0 \ ,
\end{equation}
with the equality satisfied only in the neutral, zero temperature case.
The equation of state corresponds to that of a conformal theory, and consequently the speed of sound, $c_s$, is determined by the dimension
\begin{equation}\label{eq.speedofsound}
 c_s^2  = \left( \frac{\partial P}{\partial {\cal E}}  \right)_S = \frac{1}{d-1} \ .
\end{equation}

Using the result for the internal energy \eqref{eq.ADMUV} it is immediate to prove that the first law of thermodynamics, $d{\cal E} = T\, dS + \mu\, dQ $, holds by checking
\begin{equation}\label{eq.firstlaws}
\left( \frac{ \partial {\cal E} }{ \partial S } \right)_Q = T \ , \qquad \left( \frac{ \partial {\cal E} }{ \partial Q } \right)_S = \mu \ .
\end{equation}

\vspace{15pt}

The calculation of the regularized thermodynamic potential is performed  evaluating the on-shell action \eqref{eq.model} with the Gibbons-Hawking term and counterterm included.
The leading contributions at large cutoff coming from each of the terms in the action are
\begin{subequations}
\begin{align}
- T \left( I_{adj}+I_{fun} \right) & = \frac{V_{d-1}}{2\kappa^2} \, \frac{2}{L^{d+1}}\left( r_\Lambda^d - m  + \cdots \right) \ , \\
- T I_{GH} & =  \frac{V_{d-1}}{2\kappa^2} \, \frac{d}{L^{d+1}}\left( m -  2\, r_\Lambda^d  + \cdots \right) \ , \\
- T I_{ct} & = \frac{V_{d-1}}{2\kappa^2} \,  \frac{d-1}{L^{d+1}} \left( 2\, r_\Lambda^d - m + \cdots \right) \ , 
\end{align}
\end{subequations}
where the dots stand for factors vanishing in the $r_\Lambda\to \infty$ limit.
 Once they are added up $\Omega = - T \left(I_{adj}+I_{fun}+I_{GH}+I_{ct} \right)  = - P$ follows. The thermodynamic relation
\begin{equation}
 \Omega = {\cal E} - TS - \mu\, Q \ ,
\end{equation}
also holds.

\subsection{Thermodynamic stability and extremality}

The specific heat at fixed charge is obtained straightforwardly
\begin{equation}
C_Q = T \left( \frac{\partial S}{\partial T} \right)_Q = \frac{2\pi}{\kappa^2} V_{d-1}  \frac{r_+^{d-1}}{L^{d-1}} \frac{ (d-1) \sqrt{1 + \varrho^2}}{ \sqrt{ 1 + \varrho^2} + \frac{r_+\, T_b\, L^2}{4\pi\, T\, L^2} \varrho^2} \geq 0 \ ,
\end{equation}
where  the final result is written in terms of $r_+$, $\varrho$ and $T$ to give a more compact expression and show explicitly the non-negativeness.  
At fixed chemical potential the specific heat is also non-negative definite, $C_\mu\geq0$.

The   two specific heats being non-negative, together with the positiveness of the susceptibility \eqref{eq.suscep}, are a signal that the system is thermodynamically stable.

\vspace{15pt}

To study extremality it is more adequate to use the original parameter $\rho$ instead of $\varrho$, since it is  related to the charge of the setup independently of the temperature.

At extremality a relation between the radius of the black hole, $r_+$, and the value of the parameter $\rho$ arises from equation \eqref{eq.temperature}
\begin{equation}
\rho_{ext} = \frac{r_+^{d-1}}{L^{d-1}} \sqrt{ \left( 1+\frac{d(d-1)}{T_b\, L^2} \right)^2  -1 } \ ,
\end{equation}
and physically the charge takes values only in a finite range given by $\rho\in [0,\rho_{ext}]$. 
The left limit of this range corresponds to the uncharged system and the right limit to the extremal case.
This  signals the familiar competition between gravitational attraction and electric repulsion of a charged black hole.
Black holes with large values of the charge are unstable, which is reflected in the thermodynamic analysis by a negative temperature.

The entropy density at extremality reads
\begin{equation}
S_{ext}=\frac{2\pi}{\kappa^2} V_{d-1} \frac{\rho_{ext}\, T_b\, L^2}{\sqrt{ d (d-1) } \, \sqrt{ d (d-1) + 2 \, T_b \,L^2  }} \geq 0\ ,
\end{equation}
where the equality holds only for a neutral black hole. 
The entropy density being positive for any charged configuration indicates that this solution does not describe the true ground state of the field theory.

All thermodynamic quantities  at extremality depend on $\rho_{ext}$ as dictated by dimensional analysis, with coefficient terms that depend on the dimensionless combination $T_b\, L^2$. 
For example, the chemical potential for $d>2$ reads
\begin{equation}
\mu  =  \frac{L\, \rho_{ext}^\frac{1}{d-1}}{(d-2)\lambda} \left[ \left( 1+\frac{d(d-1)}{T_b\, L^2} \right)^2  -1 \right]^\frac{d-2}{2d-2} {}_2F_1\left( \frac{1}{2} , \frac{d-2}{2d-2} ; \frac{3d-4}{2d-2} ; 1 - \left( 1+\frac{d(d-1)}{T_b\, L^2} \right)^2 \right) \ ,
\end{equation}
which is used to express physical quantities in terms of the chemical potential $\mu$ 
\begin{equation}
M \sim \mu^d \ , \qquad S_{ext} \sim \mu^{d-1} \ , \qquad \chi \sim \mu^{d-1} \ ,
\end{equation}
whereas the specific heats (both at fixed charge or chemical potential) vanish in this limit.

\subsection{Large and small charge}\label{sec.largesmallrho}

Equation \eqref{eq.smallrho} for the determination of the small $\rho$ condition  following from \eqref{eq.UVsolution} reads
\begin{equation}\label{eq.smalrhoUV}
\frac{r^{2(d-1)}}{L^{2(d-1)}} \gg \rho^2  \ .
\end{equation}
This $r$-dependent condition determines when the EMD approximation to the system \eqref{eq.model} is valid. 
The left-hand side of \eqref{eq.smalrhoUV} is bounded from below by $r_+$ and the right-hand side from above by $\rho_{ext}$.
Using these extreme values in \eqref{eq.smalrhoUV} gives a condition
\begin{equation}
 T_b\,L^2  \gg \frac{d(d-1)}{\sqrt{2}-1}   \qquad \textrm{or} \qquad  T_b\, \tilde L^2  \gg \frac{d(d-1)}{\sqrt{2}} \ .
\end{equation}
When this condition is satisfied the EMD approximation is always valid, for any value of the radius and any allowed value of the charge.
In other words, the non-linear effects are highly suppressed for large values of the tension of the brane.

\vspace{15pt}

On the other hand, the  large $\rho$ condition in the same solution is given by
\begin{equation}
1 \ll \frac{L^{2(d-1)}}{r^{2(d-1)}} \rho^2  \leq  \frac{L^{2(d-1)}}{r_+^{2(d-1)}} \rho^2 \ ,
\end{equation}
which is not satisfied in general for an arbitrary value of $\rho$ at fixed $T_b \,L^2$.
Indeed, only when
\begin{equation}
T_b\,L^2  \ll \frac{d(d-1)}{\sqrt{2}-1}   \qquad \textrm{or} \qquad  T_b\, \tilde L^2  \ll \frac{d(d-1)}{\sqrt{2}}  \ ,
\end{equation}
the near extremal ($\rho$ close to $\rho_{ext}$) solution can be described by the Nambu-Goto  approximation \eqref{eq.systemlargerho}.

\section{IR scaling solution}  \label{sec.IR}

In this section a scaling solution to the equations of motion derived from \eqref{eq.system} is studied. 
This solution arises for a simple choice of the dilatonic functions that appear in the action \eqref{eq.system}
\begin{equation}
Z_1 = e^{-\alpha\, \phi} \ , \qquad Z_2 = e^{\beta\, \phi} \ , \qquad V = 2\tilde \Lambda\,  e^{m\, \phi} \ .
\end{equation}

The metric and dilaton in this case are \cite{Pal:2012zn}
\begin{align}\label{eq.solutionPal}
ds^2 & = r^{-2\gamma} \left( - r^{2z} f(r) dt^2 + r^2 d\vec x^2 + \frac{dr^2}{r^2 f} \right) \ , \qquad f=1- \left( \frac{r_h}{r}\right)^{z + (1-\gamma)(d-1)} \ , \\
\phi & = \sqrt{2(1-\gamma) (d-1) (z-\gamma-1)} \log r \ ,
\end{align}
where the values of $d$, $z$ and $\gamma$  must be related to the parameters in the Lagrangian by
\begin{subequations}\label{eq.parameters}
\begin{align}
\alpha & = \frac{(d-1)\left(d+1+\gamma(1-d) \right)}{2 \sqrt{2(1-\gamma)(d-1)(z-\gamma-1)}} \ , \\
\beta & = \frac{d-1+\gamma(3-d)}{\sqrt{2(1-\gamma)(d-1)(z-\gamma-1)}} \ , \\
2 \tilde \Lambda & = - \left(z+(1-\gamma)(d-1)\right) \frac{ 2 (z-1) (1+\rho^2) +\rho^2 (d-1) (1-\gamma) }{\rho^2} \ , \\
m & = \frac{2 \gamma}{\sqrt{2(1-\gamma)(d-1)(z-\gamma-1)}} \ , \\
T_b & = 2(z-1) \left(z +(1-\gamma)(d-1)\right)  \frac{\sqrt{1+\rho^2}}{\rho^2}  \ .
\end{align}
\end{subequations}
The null energy condition \eqref{eq.NEC} implies that the tension of the brane is non-negative, $T_b\geq0$, and the proportionality factor of the potential is non-positive, $2\tilde \Lambda\leq0$.
The condition $T_b<2|\tilde \Lambda|$ is satisfied in the top-left and bottom-right light regions of figure \ref{fig.NEC}.
In the light region in the middle this is not guaranteed, despite the NEC being satisfied.

Deviations from the exponential behaviour in the functions $Z_1$ and $Z_2$ are interpreted as a toy model for the inclusion of quantum corrections \cite{Harrison:2012vy}. 
The class of metrics \eqref{eq.hvlmetric} has the well-known issue of having diverging tidal forces near the origin ($r=0$) at zero temperature, even when curvature invariants are finite.
This singularity can be cloaked with a black hole horizon, and therefore is considered to be mild.
In \cite{Harrison:2012vy} it was shown that generic corrections of the kind $Z_1 = e^{-\alpha \phi}+ \xi_0 + \xi_1 e^{\delta \phi} + \cdots$, which are important in the interior of the bulk geometry, cure the behaviour at the origin of the solutions in the the EMD model \cite{Harrison:2012vy,Bhattacharya:2012zu}.
The solution \eqref{eq.solutionPal} should be interpreted as a classical approximation that  suffers important IR corrections, but still be valid at intermediate values of the radial direction.

 \vspace{15pt}
 
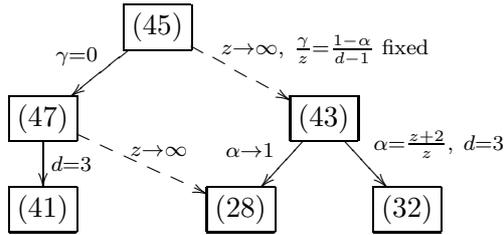
\begin{figure}[ht]
\[
\xymatrix@C=0.2cm@R=0.2cm{
 & & *+[F]{(45)}  \ar[lldd]_{\gamma=0}  \ar@{-->}[rrdd]^{\quad z\to\infty, \  \frac{\gamma}{z}=\frac{1-\alpha}{d-1} \text{ fixed}} & & &  \\
  & & & & & \\
*+[F]{(47)}  \ar[dd]^{d=3}   \ar@{-->}[ddrrr]^{z\to\infty} & & & &  *+[F]{(43)} \ar[ldd]_{\alpha\to1} \ar[ddr]^{\alpha=\frac{z+2}{z}, \ d=3} & \\
  & & & & & \\
*+[F]{(41)}  & & &  *+[F]{(28)}  & & *+[F]{(32)}
}
\]
\caption{Following the enumeration in \cite{Pal:2012zn} (inside the boxes), the top box refers to the solution written explicitly here in equations \eqref{eq.solutionPal}-\eqref{eq.parameters}, and the arrows indicate the limits that recover the rest of the IR solutions considered in \cite{Pal:2012zn}.}\label{fig.web}
\end{figure}

In  \cite{Pal:2012zn},  some limiting cases of the solution \eqref{eq.solutionPal}-\eqref{eq.parameters}  were studied. 
A scheme of the limits is given in figure \ref{fig.web}.
\begin{itemize}
\item The first limit in figure \ref{fig.web} is the $\gamma\to0$ case, i.e., the case without hyperscaling violation.
The  metric components asymptote to a Lifshitz metric with running dilaton, whose behaviour is dictated by a constant potential \cite{Taylor:2008tg}.
\item A different possibility is obtained by taking $z\to \infty$ with $\frac{\gamma}{z}$ fixed.
It is convenient to change the radial coordinate to
\begin{equation}
r = R^{1/z} \ ,
\end{equation}
and consider $R$ to be of order one.
Defining $\frac{\gamma}{z}=\frac{1-\alpha}{d-1}$, and with a rescaling  $t \to t/z$ and $\vec x \to \vec x/z$, the solution reads
\begin{equation} \label{eq.firstapprox}
ds^2 = \frac{R^{2\frac{\alpha-1}{d-1}}}{z^2} \left( -R^2 f(r) d\bar t^2 + d\vec {\bar x}^2 + \frac{dR^2}{R^2f(r)} \right) \ , \qquad f = 1 - \left( \frac{R_h}{R} \right)^{\alpha} \ .
\end{equation}
The dilaton profile and the coefficients $\alpha$, $\beta$, $m$, $2\tilde \Lambda$ and $T_b$ are obtained taking the same limit in \eqref{eq.parameters}.
In this rescaling the total charge density must be kept fixed. From   $\vec x \to \vec{x} /z$  the density $\rho$ is rescaled accordingly as  $\rho\to z^{d-1}\rho$.

The new solution is a fixed point of the equations of motion, and can be considered on its own.
 In particular, any value of $z$ is allowed despite \eqref{eq.firstapprox} being obtained  in the large $z$ limit. 
The NEC implies
\begin{equation}
\alpha > 1 \ ,
\end{equation}
independently of $z$, which is not interpreted as a dynamic exponent any longer (the geometry is not Lifshitz). 
\item As an example of a limit taken in \eqref{eq.firstapprox}, setting $\alpha=1$ the (black) $AdS_2\times R^{d-1}$ spacetime is recovered. 
\end{itemize}

\vspace{15pt}

Large and small $\rho$  (according to equations \eqref{eq.smallrho} and \eqref{eq.largerho}) is translated in the IR solution to the conditions $\rho\gg1$ or $\rho\ll1$, independently of the radial coordinate.
In the small $\rho$ case the system reduces to the Legendre transformed version of the EMD theory \eqref{eq.systemsmallrho}.
Evaluating this effective action in the IR scaling solution the potential reads
 \begin{equation}
 \tilde V \equiv V + T_b Z_1 Z_2^{(d+1)/2} = - \left[ z+(1-\gamma)(d-1)  \right] \left[ z+(1-\gamma)(d-1) -1  \right] e^{\frac{2\gamma \phi}{\sqrt{2(d-1)(1-\gamma)(z-\gamma-1)}}} \ ,
 \end{equation}
 and the dilatonic coupling in the last term of \eqref{eq.systemsmallrho} is
 \begin{equation}
\tilde Z =  Z_1^{-1}Z_2^{\frac{3-d}{2}} =  e^{\frac{2 \left( (1-\gamma)(d-1) + \gamma \right) \phi}{\sqrt{2(d-1)(1-\gamma)(z-\gamma-1)}}}\ .
\end{equation}
 These are precisely the right values to recover the solutions studied in   \cite{Gouteraux:2011ce,Taylor:2008tg}.
On the contrary, if $\rho$ is large,  action \eqref{eq.systemlargerho} is recovered.
The explicit coefficients in this limit are\footnote{ $\alpha$ is not provided since $Z_1$ is not present in the effective action at large $\rho$.}
\begin{subequations}
\begin{align}
\beta & = \frac{d-1+\gamma(3-d)}{\sqrt{2(1-\gamma)(d-1)(z-\gamma-1)}} \ , \\
2 \tilde \Lambda & = - \left(z+(1-\gamma)(d-1)\right) \left( 2 (z-1) + (d-1) (1-\gamma) \right) \ , \\
m & = \frac{2 \gamma}{\sqrt{2(1-\gamma)(d-1)(z-\gamma-1)}} \ , \\
\tilde T_b & =   2(z-1) \left(z +(1-\gamma)(d-1)\right)  \ .
\end{align}
\end{subequations}

\subsection{Thermodynamics of the IR scaling solution}

The temperature of the field theory is canonically identified with the Hawking temperature, and the entropy density with the Bekenstein-Hawking entropy
\begin{equation}\label{eq.LifshitzT}
T = \frac{r_h^z}{4\pi} \left( z +(d-1)(1-\gamma) \right) \ , \qquad
S = \frac{2\pi}{\kappa^2} V_{d-1} \left( \frac{4 \pi \, T}{z+(d-1)(1-\gamma)} \right)^{\frac{(d-1)(1-\gamma)}{z}} \ .
\end{equation}
Notice that  the temperature  being non-negative requires $z+(d-1)(1-\gamma)\geq0$, which from the NEC condition \eqref{eq.NEC} imposes $z\geq1$.

The specific heat reads
\begin{equation}\label{eq.IRcv}
C_V= T \partial_T S = \frac{2\pi}{\kappa^2} V_{d-1}  \frac{(d-1)(1-\gamma)}{z} \left( \frac{4 \pi \,T}{z+(d-1)(1-\gamma)} \right)^{\frac{(d-1)(1-\gamma)}{z}} \ ,
\end{equation}
and is non-negative if $1-\gamma\geq0$.

\vspace{15pt}

The calculation of the internal and free energies of the setup requires the use of a holographic normalisation  for Lifshitz and hyperscaling violating spacetimes.
The set of counterterms that should be added depends on the dimensionality, the dynamical exponent and the hyperscaling violation coefficient.
For the Einstein-DBI-Dilaton type of action discussed in this paper no results are available so far.
A different possibility, followed here, to regularise the energies is to use instead background subtraction: calculate the difference in the internal and free energies between the euclidean version of the black hole  and the euclidean version of the horizonless solution, with the euclidean time periodicity identified to describe the same temperature.

The ADM mass of the black hole becomes the holographic dual to the  internal energy of the field theory.
The  definition of the ADM mass is given in terms of the trace of the extrinsic curvature for a radial slice of a constant-$t$ surface, $\Theta=\frac{1}{\sqrt{g_{rr}}}\partial_r \log \sqrt{g_{xx}^{d-1}}$ evaluated at a large cutoff
 \begin{equation}
M_{r_h} = -  \frac{1}{\kappa^2} \sqrt{|g_{tt}|} \int d^{d-1}x \sqrt{g_{xx}^{d-1}} \, \Theta \Big|_{r=r_\Lambda} \ ,
 \end{equation}
with the solution describing a field theory at temperature $T$ from \eqref{eq.LifshitzT}.
The regularized quantity 
\begin{equation}
M=\lim_{r_\Lambda\to\infty}\left( M_{r_h}- \sqrt{f} \, M_{r_h=0} \right) \ ,
\end{equation}
corresponds to the regularized internal energy.
The  factor  $\sqrt{f}$ is included to ensure that the geometries being compared have the same euclidean time periodicity.
 The final result is
 \begin{equation}\label{eq.admmass}
M = \frac{V_{d-1}}{2\kappa^2} (d-1)(1-\gamma)  \left( \frac{4 \pi \,T}{z+(d-1)(1-\gamma)} \right)^{1+ \frac{(d-1)(1-\gamma)}{z}}\ ,
\end{equation}
from where $dM=TdS$ follows immediately.

The calculation of the regularized thermodynamic potential is performed by evaluating on-shell action with the Gibbons-Hawking term included.
One difference to the UV-complete solution of section \ref{sec.UV} is that now the charge of the $U(1)$ field is fixed to support the Lifshitz, hyperscaling violating geometry. 
In other words,the normalizable mode of the gauge field is kept constant, working with Neumann boundary conditions, and the action to consider is  \eqref{eq.system}.
This is equivalent to working in the canonical ensemble.

After background subtraction the regularized on-shell action \eqref{eq.system} is
 \begin{equation}
- T I_E = \frac{1}{2\kappa^2} V_{d-1} (\gamma-z)\, r_h^{z+(d-1)(1-\gamma)}  \ ,
 \end{equation}
 where the  $T=0$ term in the subtraction is corrected with a $\sqrt{f}$ factor to ensure the   euclidean geometries being compared have the same periodicity of the euclidean time circle.
The Gibbons-Hawking term gives a finite contribution to the on-shell evaluation of the action, proportional to the hyperscaling violation parameter
  \begin{equation}
- T  I_{GH} =  -\frac{1}{2\kappa^2} V_{d-1} \gamma \, r_h^{z+(d-1)(1-\gamma)} \ .
 \end{equation}
Adding up these two contributions the total free energy in the canonical ensemble is
\begin{equation}
F = - T I_E - T I_{GH}  =  M - TS = -z\, \frac{V_{d-1}}{2\kappa^2}   \left( \frac{4 \pi T}{z+(d-1)(1-\gamma)} \right)^{1+ \frac{(d-1)(1-\gamma)}{z}}\ ,
\end{equation}
and the negativity shows that for any value of the temperature these solutions are preferred over the corresponding thermal solutions.

The speed of sound for $z\neq 1$ cannot be calculated.
 The reason is that the speed of sound is obtained from the adiabatic variation of the pressure with the matter density.
 When $z=1$ the system is relativistic and  the matter and energy densities can be identified, but for $z>1$ this identification is not possible \cite{HoyosBadajoz:2010kd}.

\section{Fluctuations and transport} \label{sec.fluctuations}

In sections \ref{sec.UV} and \ref{sec.IR} two solutions to the system \eqref{eq.model} were described. 
It was shown that these solutions do not show thermodynamic instabilities.
In this section the aim is to study whether the solutions are perturbatively stable under fluctuations of the fields
\begin{subequations}
\begin{align}
g_{\mu\nu} & \to g_{\mu\nu}(r) + h_{\mu\nu}(x^0,x^1,r) \ , \\
 \phi & \to \phi(r) + \varphi(x^0,x^1,r) \ , \\
  A  & \to A_t(r) dt + a_\mu(x^0,x^1,r) dx^\mu \ , \label{eq.efieldfluc}
\end{align}
\end{subequations}
with $x^d=r$ the radial coordinate.
The background SO($d-1$) symmetry is used to align the fluctuation momentum, $q$, along the $x^1$ direction.
Furthermore, these fluctuations allow to extract holographically the two-point functions of the field theory.
From the two-point functions different transport coefficients like the conductivity or the diffusion of the charge constant can be extracted.

\vspace{15 pt}

The fluctuations are split into independent channels according to their transformation properties under the surviving little group SO($d-2$) (for $d>2$).
The different modes fall in one of the following three categories: modes transforming as  scalars under SO($d-2$) (from now on scalar modes), modes transforming as  vectors components (vectorial modes) and finally modes transforming as tensors components (tensorial modes).

In addition to this group theory decomposition, invariance under diffeomorphism and gauge transformations
\begin{subequations}\label{eq.gaugediffeo}
\begin{align}
h_{\mu\nu} & \to h_{\mu\nu} - \nabla_\mu \, \xi_\nu - \nabla_\nu \, \xi_\mu \ , \\
a_\mu & \to a_\mu - \xi^\nu \nabla_\nu \, A_\mu - A^\nu \nabla_\mu \, \xi_\nu  - \nabla_\mu \, \chi \ , \\
\varphi & \to \varphi - \xi^\mu \nabla_\mu \, \phi \ ,
\end{align}
\end{subequations}
is used to identify the  independent physical modes.
In these equations   $\nabla_\mu$ is the covariant derivative with respect to the background metric and the components of $\xi_\mu$ and $\chi$ depend on $x^0$, $x^1$ and $r$.

The tensorial modes are automatically invariant under gauge and diffeomorphism transformations.
This is not the case for perturbations in the vectorial and scalar channels, which need to be combined into invariant combinations.
To write these it is convenient to work in the radial gauge, $h_{\mu r}=a_r=0$, and  define the quantities
\begin{equation}
H_{00}= g_{tt} h_{00} \ , \quad H_{11} = g_{xx} h_{11} \ ,  \quad H_{\mu i} = g_{xx} h_{\mu i} \ ,  \quad H_{01} = g_{xx} h_{01} \ ,
\end{equation}
where in this equation and in the rest of this paper $g_{tt}=|g_{tt}|$ for notational simplicity, and $i,j=\{2,\cdots,d-1\}$.

\paragraph{Tensorial modes:}

For $d>3$ there are tensorial modes consisting of the $h_{ij}$ perturbations.
Using the Fourier decomposition 
\begin{equation}
H_{ij} = e^{-i \left( \omega \, x^0 + q \, x^1 \right)} L_{H,2}(r) \ ,
\end{equation}
the equation of motion reads
\begin{equation}
L_{H,2}''+ \partial_r \log \left[  \sqrt{ \frac{ g_{tt} \, g_{xx}^{d-1} }{ g_{rr} } }  \right] L_{H,2}' + g_{rr} \left( \frac{ \omega^2 }{ g_{tt} } - \frac{ q^2 }{g_{xx} } \right)  L_{H,2} = 0  \ ,
\end{equation}
which is simply the equation for a massless scalar in a given background spacetime.

\paragraph{Vectorial modes:}

Vectorial modes exist for $d>2$. 
The equations of motion couple the fields $a_i$ and $h_{0i}$, $h_{1i}$, $h_{ri}$ for $i=2,\cdots, d-1$. 
Although $a_i$ are gauge and diffeomorphism invariant, this is not the case for the metric perturbations.
Writing the Fourier decomposition of the independent modes, it is easy to check that the combination $q \, H_{0i} + \omega \, H_{1i} $ does satisfy gauge and diffeomorphism invariance.

Defining
\begin{subequations} \begin{align}
\lambda \, a_i & = e^{-i \left( \omega \, x^0 + q \, x^1 \right)} L_{a,1}(r) \ , \\
 q \, H_{0i} + \omega \, H_{1i} & = e^{-i \left( \omega \, x^0 + q \, x^1 \right)} L_{H,1}(r) \ ,
\end{align}\end{subequations}
the equations  of motion read
\begin{subequations} \begin{align}
& L_{a,1}'' + \partial_r \log \left[  \frac{\sqrt{ g_{tt} \left(  \rho^2 + Z_1^{2}Z_2^{d-1} g_{xx}^{d-1} \right) }}{ \sqrt{g_{rr}} \, g_{xx}\,Z_2 }  \right] L_{a,1}'   \nonumber \\ 
& \quad +  g_{rr} \left(  \frac{\omega^2 }{ g_{tt} } - \frac{  Z_1^{2} Z_2^{d-1} g_{xx}^{d-1} }{ \rho^2 + Z_1^{2} Z_2^{d-1} g_{xx}^{d-1} } \frac{ q^2 }{ g_{xx }} - \frac{  Z_2  T_b\, \rho^2\,  \omega^2}{ g_{xx}^\frac{d-3}{2} (\omega^2 g_{xx}-q^2 g_{tt}) \sqrt{  \rho^2 + Z_1^{2}Z_2^{d-1} g_{xx}^{d-1} } } \right) L_{a,1}  \nonumber \\
& \quad - \sqrt{ \frac{ g_{rr} \, g_{tt} }{  \rho^2 + Z_1^{2}Z_2^{d-1} g_{xx}^{d-1} } } \, \frac{ Z_2 q \, \rho \, g_{xx} }{ (q^2 g_{tt} - \omega^2 g_{xx})  }  L_{H,1}' = 0\ , \label{eq.vectorfluc}\\
& L_{H,1}'' +  \partial_r  \log \left[  \sqrt{ \frac{ g_{tt} \, g_{xx}^{d+1} }{ g_{rr} } } \, \frac{ 1 }{ q^2 g_{tt} - \omega^2 g_{xx} }  \right]  L_{H,1}'  + g_{rr} \left(  \frac{\omega^2 }{ g_{tt} } -  \frac{ q^2 }{ g_{xx }} \right) L_{H,1} \nonumber \\
& \quad + \frac{ q\, T_b\, \rho\, \sqrt{ g_{rr} \, g_{tt} } }{ g_{xx}^\frac{d+1}{2}  } \,  L_{a,1}' + \sqrt{ \frac{ g_{rr} }{ g_{tt} } } \, \frac{ q\, T_b\, \rho\, \omega^2 (g_{tt} \, g_{xx}'-g_{xx} \, g_{tt}') }{ g_{xx}^\frac{d+1}{2}  (q^2 g_{tt} - \omega^2 g_{xx}) } L_{a,1} = 0 \quad \ .\label{eq.metricfluc}
 \end{align}\end{subequations}
Equation \eqref{eq.metricfluc} decouples in the three limits $T_b=0$, $\rho=0$ and $q=0$, recovering  the metric fluctuation equation of motion written in  \cite{Mas:2007ng} for the vectorial channel.\footnote{
In the $q\to 0$ limit the two equations decouple and equation \eqref{eq.vectorfluc} should  become   eq.(104) in \cite{Pal:2012zn}, but in the last term multiplying $L_{a,1}$ there is relative $- 1$ which in that paper reads $+2$. 
This relative factor matters, since this term is the responsible for the strength of the delta function at zero frequency in the optical conductivity of the system.
}

\paragraph{Scalar modes:}

 The rest of the fluctuations organise themselves into three diffeomorphism and gauge invariant scalars under SO($d-2$), given in Fourier modes by (no sum on $H_{ii}$)
\begin{subequations} \begin{align}
\varphi - \phi' \frac{ g_{xx} }{ g_{xx}' } H_{ii} & = e^{-i \left( \omega \, x^0 + q \, x^1 \right)} L_{\varphi,0}(r) \ , \\
\lambda \left( q \, a_0 + \omega \, a_1 - q \, A_t' \, \frac{ g_{xx} }{ g_{xx}' }  H_{ii} \right) & = e^{-i \left( \omega \, x^0 + q \, x^1 \right)} L_{a,0}(r) \ , \\
- q^2 \frac{g_{tt}}{g_{xx}} H_{00} + \omega^2 H_{11} + 2\, q \, \omega \, H_{01} + \left( \frac{g_{tt}'}{g_{xx}'}\, q^2-\omega^2 \right) H_{ii} & = e^{-i \left( \omega \, x^0 + q \, x^1 \right)} L_{H,0} (r) \ .
 \end{align}\end{subequations}
The coupled system of equations of motion for $L_{\varphi,0}$, $L_{a,0}$and $L_{H,0}$ is not written explicitly for clearness.
The equations of motion for the non-invariant quantities are provided in appendix \ref{app.equations}, as well as (non-independent) first order constraints.
From these, some limits can be taken. 
When $T_b=0$ the perturbation $L_{a,0}$ decouples from the other two, which becomes the system described  in \cite{Mas:2007ng}.\footnote{Compare equations \eqref{eq.app.scal1}-\eqref{eq.app.scal2}, \eqref{eq.app.scal5} and \eqref{eq.app.scal7}-\eqref{eq.app.scal8} to the equations in \cite{Mas:2007ng}.}
In the limit of zero momentum the longitudinal mode of the gauge field is not defined, and the system of equations in \eqref{eq.app.scal4} and \eqref{eq.app.scal7b} gives the $q=0$ vector fluctuation in \eqref{eq.vectorfluc}.

\subsection{Conductivity}

The optical conductivity is obtained from the retarded current-current correlator at zero momentum, via the relation
\begin{equation}
\sigma (\omega) =   \frac{-i}{\omega} G_{JJ}^{(R)}(\omega,q=0) \ .
\end{equation}
Holographically, the current-current correlator is obtained from the transverse fluctuation of the electric field \eqref{eq.efieldfluc}.
As commented above, at zero momentum the equation for the fluctuations of the electric field \eqref{eq.vectorfluc} decouples, and the quadratic action from which this fluctuation is derived is
\begin{align}\label{eq.conducaction}
S_{aa} = - \frac{1}{2}  \int dr \frac{d\omega \,d^{d-1} q}{(2\pi)^d}\sqrt{ \frac{ g_{tt} }{ g_{rr} } } \Sigma(r) \left[ L_{a,1}'^2  - g_{rr} \left(  \frac{\omega^2 }{ g_{tt} }  - \Pi(r) \right) L_{a,1}^2   \right] \ ,
\end{align}
where
\begin{equation}
\Sigma(r) =  \frac{ T_b}{2\kappa^2} \frac{\sqrt{  \rho^2 + Z_1^{2}Z_2^{d-1} g_{xx}^{d-1} } }{ g_{xx}\,Z_2 }  \ , \quad \Pi(r) =  \frac{  Z_2  T_b\, \rho^2\,  g_{xx}^\frac{1-d}{2}  }{ \sqrt{  \rho^2 + Z_1^{2}Z_2^{d-1} g_{xx}^{d-1} } } \ .
\end{equation}
Notice that both $\Sigma(r)$ and $\Pi(r)$ are finite at the horizon.
In this section  finite temperature is assumed, so near the horizon $g_{tt} \sim (r-r_h)$ and $g_{rr}\sim (r-r_h)^{-1}$.

The standard prescription to calculate the retarded Minkowskian correlator \cite{Policastro:2002se} consists of first imposing an ingoing-wave condition for $L_{a,1}$ near the horizon. 
After fixing an overall normalisation an integration of the equation of motion \eqref{eq.vectorfluc} is performed, and  the normalizable and non-normalizable modes (call them $\ell_1$ and $\ell_0$  respectively) are obtained from the asymptotic expansion of the fluctuation $L_{a,1}$. 
For example, in the UV-complete solution, for $d=3$
\begin{equation}\label{eq.canonicalform}
L_{a,1} = \ell_0 + \frac{\ell_1}{r} + \cdots \ .
\end{equation}
Finally, the retarded current-current correlator is proportional to their ratio $G_{JJ}^{(R)} \sim \ell_1/\ell_0$.

\vspace{15pt}

In this paper a strategy based on the one  described in  \cite{Iqbal:2008by} is employed. 
This provides a different (equivalent) way to calculate the field theory conductivity.
 The main difference with the approach in  \cite{Iqbal:2008by} is that here  a non-trivial mass term in the action, $\Pi(r)$, is allowed.
 This term is responsible for two effects: (1) it determines the strength of the delta function at zero momentum in the conductivity, and (2) the finite part of the radial-dependent DC conductivity (to be defined shortly) runs with the radius.

To see how these effects arise, define the frequency-dependent conductivity function, $\sigma_\omega(r)$, whose value at the boundary coincides with the value of the optical conductivity in the field theory, $\sigma_\omega(r\to\infty) = \sigma(\omega)$.
To find the equation of motion governing this function  first write the canonical momentum associated to the fluctuation\footnote{Actually,  the fluctuation of the gauge field $a_1=L_{a,1}/\lambda$ must be considered. Instead, $L_{a,1}$ is used here, and $\lambda$ reinstated at the end.} $L_{a,1}$,  referred to as the current function
\begin{subequations}
\begin{align}
J(r) &= \frac{\delta S_{aa}}{ \delta L_{a,1}'} = - \sqrt{ \frac{ g_{tt} }{ g_{rr} } } \Sigma(r) L_{a,1}' \ , \\
 J'(r) & = - \frac{\delta S_{aa}}{\delta L_{a,1}} = -  \sqrt{ \frac{ g_{tt} }{ g_{rr} } } \Sigma(r)  g_{rr}\left(  \frac{\omega^2 }{ g_{tt} }  - \Pi(r) \right) L_{a,1} \ . \label{eq.Jprime}
\end{align}
\end{subequations}
Now invoke Ohm's law $J_i=\sigma_{ij} E_j$, and assume that it holds for the current and conductivity functions
\begin{equation}
J(r) = \sigma_\omega(r) E(r) = - \sigma_\omega(r) i \omega L_{a,1} \ ,
\end{equation}
where  the Fourier decomposition $\lambda\, a_{1}(x^0,r)=e^{-i \, \omega \, x^0} L_{a,1}(r)$ has been used.
After a derivative and some manipulations this expression leads to
\begin{equation}\label{eq.iqliuconduc}
\sigma_\omega'(r) =  - \frac{i}{\omega} \sqrt{ \frac{ g_{rr} }{ g_{tt} } } \left[  \omega^2\left( \Sigma(r) -  \frac{\sigma_\omega(r)^2}{ \Sigma(r) }     \right)  - g_{tt}\,  \Sigma(r) \,  \Pi(r)  \right] \ .
\end{equation}

When $\Pi(r)=0$ the derivative of the conductivity function is proportional to the frequency.
This in turn means that the DC conductivity is constant in the radial variable $\sigma_{\omega\to0}(r)=\sigma_{DC}$, and imposing regularity at the horizon gives $\sigma_{DC}=\Sigma(r_h)$ \cite{Iqbal:2008by}. 
For non-vanishing mass term $\Pi(r)\neq0$, inspection of equation \eqref{eq.iqliuconduc} suggests a low frequency expansion of the form\begin{equation}
\sigma_\omega(r) = \frac{\sigma_{-1}(r)}{\omega}+\sigma_0(r) + \sigma_1(r) \omega + \cdots \ , 
\end{equation}
which in turn gives the following equations order by order in the frequency
\begin{subequations}\begin{align}\label{eq.imaginarypole}
\sigma_{-1} ' & = i\, \sqrt{ \frac{ g_{rr} }{ g_{tt} } }  \left( g_{tt}\,\Sigma \, \Pi + \frac{\sigma_{-1}^2}{\Sigma} \right) \ , \\ \label{eq.DCconduc}
\sigma_{0} ' & = i\, \sqrt{ \frac{ g_{rr} }{ g_{tt} } }  \, \frac{2 \, \sigma_0 \, \sigma_{-1}}{\Sigma} \ , \\ \label{eq.firstcorrec}
\sigma_{1} ' & = i\, \sqrt{ \frac{ g_{rr} }{ g_{tt} } } \left(  \frac{\sigma_0^2+2 \, \sigma_{-1} \, \sigma_1}{\Sigma} - \Sigma \right) \ , \\
& \ \vdots \nonumber
\end{align}\end{subequations}

The mass term is responsible for a pole at $\omega=0$ in the conductivity function.
From \eqref{eq.imaginarypole} $\sigma_{-1}$ is purely imaginary. 
Asymptotically this means that the optical conductivity presents an imaginary pole at zero frequency. 
By means of the Kramers-Kronig relations this translates into a delta peak at $\omega=0$ in the real part.
This delta peak is a consequence of translational invariance: momentum is conserved in the system and therefore perfect conduction at zero momentum is expected.

Integration of equation \eqref{eq.imaginarypole} has to be done numerically, in general. 
The condition to impose is regularity near the horizon, which implies 
$
\sigma_{-1} =  {\cal O}(r-r_h)  
$.
After the integration is performed  the pole in the imaginary part of the optical conductivity is obtained from the asymptotic constant to which $\sigma_{-1}$ flows at large radius.
With the solution to \eqref{eq.imaginarypole} at hand, the $\omega\to0$ limit of the real part of the optical conductivity is obtained integrating \eqref{eq.DCconduc}
\begin{equation}
\sigma_0(r) = k_0 \, \exp \left[ 2i\,\int^r \sqrt{\frac{g_{rr}(r')}{g_{tt}(r')}}  \,  \frac{\sigma_{-1}(r')}{\Sigma(r')} \, dr' \right] \ ,
\end{equation}
with $k_0$ an integration constant.
To fix its value  plug this solution into \eqref{eq.firstcorrec}. 
Since $\sigma_0(r)$ is finite at the horizon, regularity of $\sigma_1(r)$ implies that at the horizon $\sigma_{0}(r_h)=\Sigma(r_h)$. 
Therefore
\begin{equation}\label{eq.myILeq}
\sigma_0(r) = \Sigma(r_h) \exp \left[2i\, \int^r_{r_h} \sqrt{ \frac{g_{rr}(r')}{g_{tt}(r')} } \, \frac{\sigma_{-1}(r')}{\Sigma(r')} \, dr' \right] \ ,
\end{equation}
and the finite part of the zero-frequency conductivity is given by $\sigma_{DC}=\sigma_0(r\to\infty)$. 

This equation generalises the result in \cite{Iqbal:2008by} to the case where there is a mass term in the equation of motion for the fluctuation, which is responsible of the presence of a zero-frequency pole in the imaginary part of the optical conductivity and the running of the $\omega\to0$ limit of the real part.
Notice that $\sigma_0(r)$ is real since $\sigma_{-1}(r)$ is imaginary.
Generically $\sigma_{even}(r)$ are real functions and $\sigma_{odd}(r)$ are purely imaginary, reflecting time reversal symmetry of the system.

\vspace{15pt}

In principle counterterms must be added to the on-shell action of the gauge field fluctuations.
These have to be given by quadratic, covariant, gauge-invariant expressions (up to the presence of logarithms of the cutoff \cite{Karch:2007pd} when $d=4$) of the vector field.
The simplest term to write down is $\int d^dx \sqrt{-\gamma} \, F_{ab}F^{ab}$, and contributes to the $\delta A= e^{-i \, \omega \,x^0}a_x (r) dx$ fluctuation with a factor proportional to $\omega^{2}\,  r_\Lambda^{d-4}$ for the solution in section \ref{sec.UV}.
For the $d=3$ case this counterterm does not survive the large cutoff limit, and for $d=4$ the cutoff dependence is $\log[r_\Lambda]$.
This factor in the on-shell action modifies equation \eqref{eq.Jprime} with a term that goes like $\omega^2 r_{\Lambda}^{d-4} L_{a,1}$.
In \eqref{eq.iqliuconduc} this appears inside the $\omega^2$ parenthesis as $\sqrt{\tfrac{g_{tt}}{g_{rr}}}\,  r_{\Lambda}^{d-4}$, and is propagated into equation \eqref{eq.firstcorrec}.
However, at the horizon this contribution vanishes, and the rest of the discussion still holds.
Only when calculating the AC conductivity at $d>3$ the contribution from the counterterms is needed to cancel divergences.

\subsubsection{DC conductivity of the UV-complete solution}

From now on only the case $d=3$ is considered for the sake of concreteness, and rescaling of the Minkowski and radius coordinates is used  to set $L=r_+=1$ without loss of generality. 
The temperature and chemical potential are determined  in terms of the dimensionless constants $T_b \, L^2$ and $\rho$, and the physical results are functions of the  dimensionless ratio\footnote{To avoid writing an explicit factor of $\lambda$ in the dimensionless ratio $\tfrac{T}{\lambda\, \mu}$ from \eqref{eq.mu}, this factor is reabsorbed in the chemical potential from now on.} $\tfrac{T}{\mu}$. 
For a fixed theory (fixed value of $T_b \, L^2$) exploring the range $\rho\in[0,\rho_{ext}]$ is equivalent to explore the range $\tfrac{T}{\mu} \in [0,\infty)$.

For the UV complete solution, the asymptotic solution for $\sigma_{-1}$ near the horizon from equation
\eqref{eq.imaginarypole} is
\begin{equation}
\sigma_{-1}(r) \approx i\,\frac{T_b^2}{2\kappa^2}\, \rho^2 (r-1) \left( 1 - \left( 2+ \frac{T_b \rho^2}{\sqrt{1+\rho^2} (6+T_b-T_b\sqrt{1+\rho^2}) } \right) (r-1) + \cdots \right) \ .
\end{equation}
Near the boundary the solution to the differential equation is
\begin{equation}
\sigma_{-1} (r) \approx \varsigma_{-1} - i \frac{\varsigma_{-1}^2}{r} - \frac{\varsigma_{-1}^3}{r^2} + \cdots \ ,
\end{equation}
where $\varsigma_{-1}$ is identified with the residue of the pole at $\omega=0$ in the optical conductivity.
\begin{figure}[tb]
\begin{center}
\includegraphics[scale=0.6]{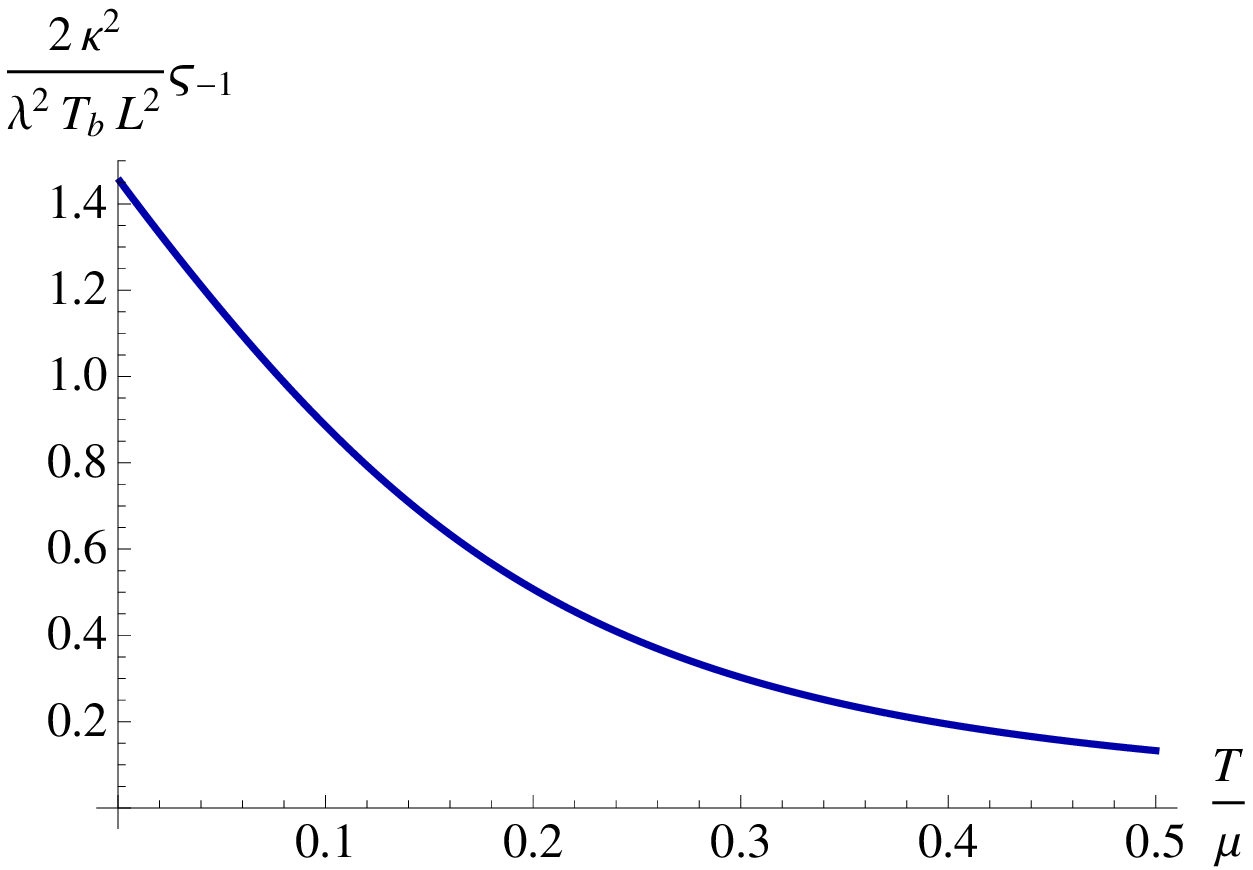}
\includegraphics[scale=0.6]{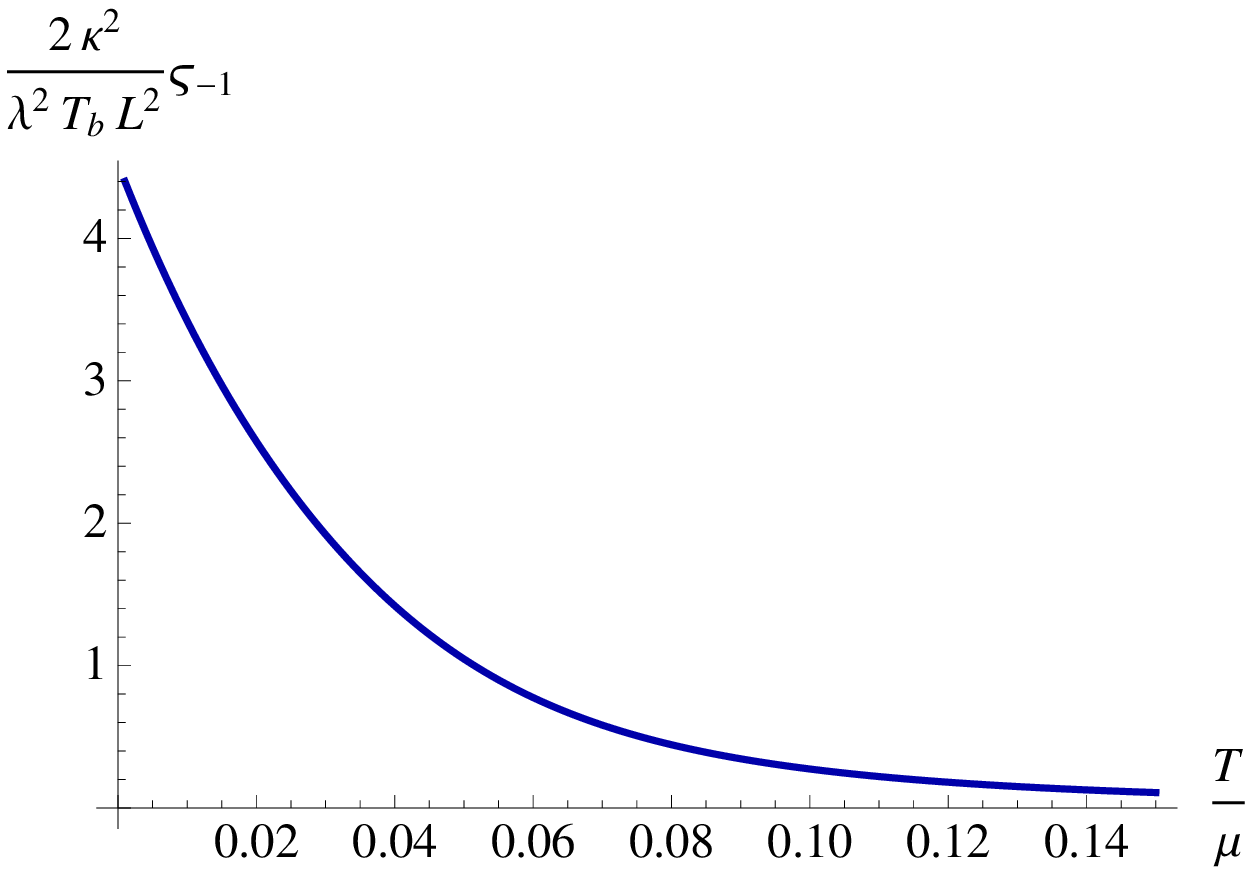}
\caption{Residue of the pole at zero frequency in the imaginary part of optical conductivity for $T_b \, L^2=2$ (left) and $T_b \, L^2=0.1$ (right). The curve coincides with the hydrodynamic result \eqref{eq.impolehydro}.}\label{fig.UVimaginaryconduc}
\end{center}
\end{figure}
The result of this calculation is shown in figure \ref{fig.UVimaginaryconduc} for $T_b \, L^2=2$ and  $T_b \, L^2=0.1$.
The numeric result in this figure coincides, within numeric accuracy, with the prediction from hydrodynamics \cite{Hartnoll:2007ih}
\begin{equation}\label{eq.impolehydro}
\varsigma_{-1} = \frac{1}{V_{2}}\frac{Q^2}{{\cal E}+P} \ ,
\end{equation}
where the value is corrected by the volume of $\mathbb{R}^{2}$.

The radial profile $\sigma_{-1}(r)$ is used to obtain  $\sigma_0(r)$ from \eqref{eq.myILeq}. 
The value at the horizon is given, for generic dimension, by  
\begin{equation}\label{eq.horizonDC}
\sigma_{0}(r_+) = \frac{\lambda^2\, T_b \, L^2}{2\kappa^2} \, \frac{\sqrt{\rho^2+ \frac{r_+^{2(d-1)}}{L^{2(d-1)}}}}{r_+^2} \ .
\end{equation}
\begin{figure}[tb]
\begin{center}
\includegraphics[scale=0.6]{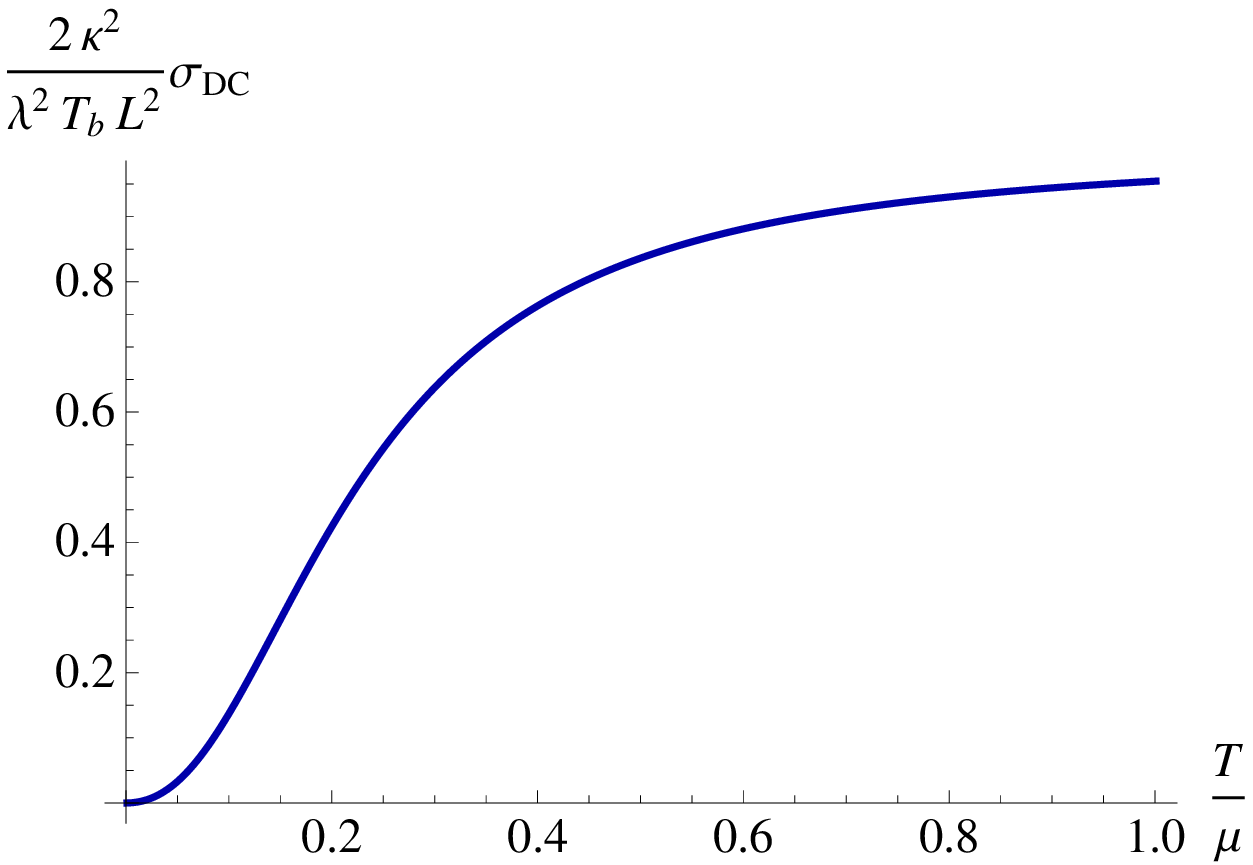}
\includegraphics[scale=0.6]{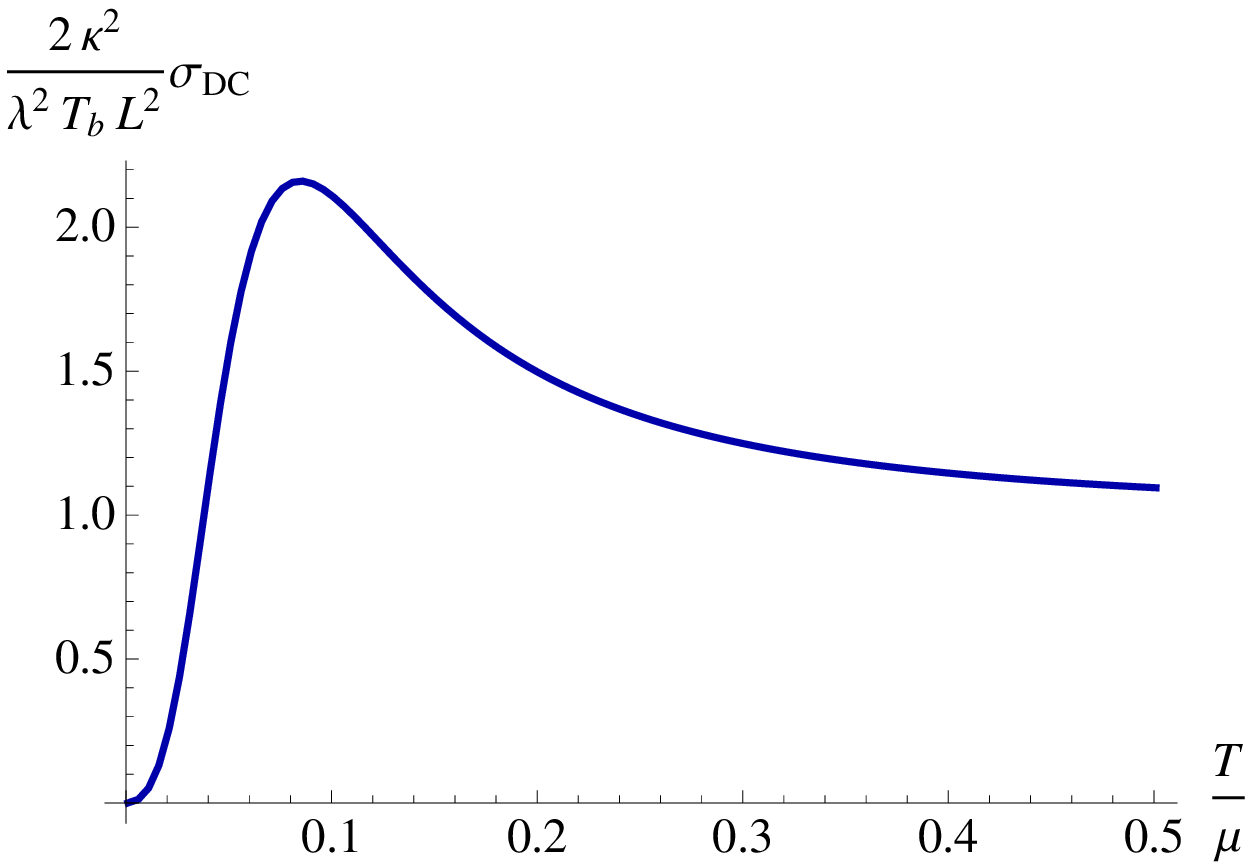}
\caption{Value of the DC optical conductivity for $T_b \, L^2=2$ (left) and $T_b \, L^2=0.1$ (right).}\label{fig.UVDCconduc}
\end{center}
\end{figure}
The running of $\sigma_0$ interpolates between this result and the actual  value of the optical DC conductivity of the field theory, for which two examples are given in figure \ref{fig.UVDCconduc}. These calculations for $d=3$ coincide with the expression \cite{Hartnoll:2007ip}
\begin{equation}
\sigma_{DC} = \sigma_0(r_+) \left( \frac{S \, T}{{\cal E}+P} \right)^2 \ .
\end{equation}

\vspace{15pt}

For any value of the temperature, a pole in the imaginary part of the conductivity is observed in figure \ref{fig.UVimaginaryconduc}.
This pole implies the existence of a delta function at zero frequency in the real part of the optical conductivity.
This is an effect of conservation of momentum: an induced current in a system with no dissipation of momentum  persists for an infinite time, which is reflected in perfect conduction.
The amplitude $\varsigma_{-1}$ is larger at lower values of the temperature (or equivalently, large values of the chemical potential), and vanishes at infinite temperature (vanishing $\mu$).

Focusing now on figure \ref{fig.UVDCconduc},  $\sigma_{DC}=\tfrac{\lambda^2 T_b\, L^2}{2\kappa^2}$ when $T\gg \mu$, and  vanishes in the opposite limit. 
On the left-hand side graph, for $T_b \, L^2=2$, the curve is monotonic. 
The behaviour is qualitatively  similar to the one observed in the Reissner-Nordstr\"om case \cite{Hartnoll:2009sz}, with vanishing DC conductivity at zero temperature and approaching a constant at large temperatures.\footnote{Compare to  figure 6 in  \cite{Hartnoll:2009sz}. On the left-hand side graph there are three plots of the (frequency-dependent) optical conductivity as a function of $\tfrac{\omega}{T}$ for different values of $\tfrac{T}{\mu}$. Focusing on the $\omega\to0$ limit of those plots it is observed that for larger $\tfrac{T}{\mu}$, $\sigma_{DC}$ is closer to $1$ (but slightly below that value), and vice versa, for small $\tfrac{T}{\mu}$ the value of the DC limit approaches zero.}
Increasing the value of $T_b \, L^2$ does not change this monotonicity, as is expected since for large values of the tension of the brane the EMD system is recovered.

On the right-hand side graph of \ref{fig.UVDCconduc} the value of the tension of the brane is lowered and the monotonic behaviour of $\sigma_{DC}$ with $\tfrac{T}{\mu}$ is lost.
Now the DC conductivity presents a maximum.
When $T_b \, L^2$ is lowered the value of this maximum increases, and its position is displaced to smaller values of $\tfrac{T}{\mu}$. 
Notice that there is no salient feature in the plot for the pole in figure \ref{fig.UVimaginaryconduc} at the corresponding value for $T_b \, L^2$.
For large values of $\tfrac{T}{\mu}$, to the right of the maximum, the value of $\sigma_{DC}$ is well approximated by the conductivity at the horizon \eqref{eq.horizonDC}. 
The reason is that the mass term vanishes in the $\rho\to 0$ limit, implying that the DC conductivity does not run  \cite{Iqbal:2008by}. 
This limit is the dilute limit in which the thermal contributions dominate, and equivalently $\tfrac{S\, T}{{\cal E}+P}\to 1$ when $\mu\to0$.
For low $\tfrac{T}{\mu}$ backreaction of the DBI term into the gravity is non-negligible and the DC conductivity falls to zero when $\tfrac{T}{\mu}\to0$.
\begin{figure}[tb]
\begin{center}
\includegraphics[scale=0.8]{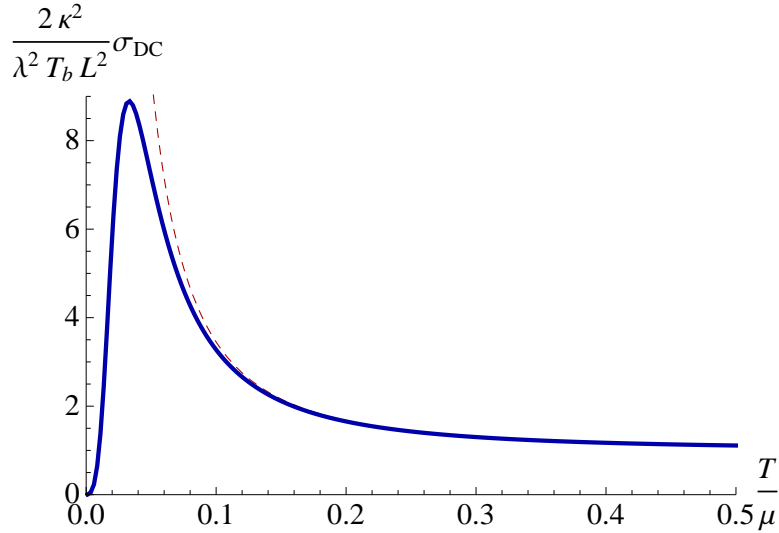}
\caption{Value of the DC  optical conductivity for $T_b \, L^2=0.01$ (straight line) and value at the horizon (dashed thin line).}\label{fig.UVwrapup}
\end{center}
\end{figure}
In figure \ref{fig.UVwrapup} the behaviour of the DC conductivity as a function of $\tfrac{T}{\mu}$ for  $T_b \, L^2=10^{-2}$ is shown, with the value at the horizon superimposed.

The value \eqref{eq.horizonDC} coincides with the value obtained for a probe brane in the background \eqref{eq.UVsolution}, following the calculation in \cite{Karch:2007pd}.
Notice that this background knows about the value of the chemical potential via the charge of the black hole.
Taking a charged probe in the neutral background misses part of the contribution coming from the larger horizon in the charged case, which at low values of the chemical potential is a negligible effect, but for temperatures of the order of the chemical potential has drastic consequences, as seen in figures \ref{fig.UVDCconduc} or \ref{fig.UVwrapup}.

\subsubsection{AC conductivity of the UV-complete solution}

\begin{figure}[tb]
\begin{center}
\includegraphics[scale=0.55]{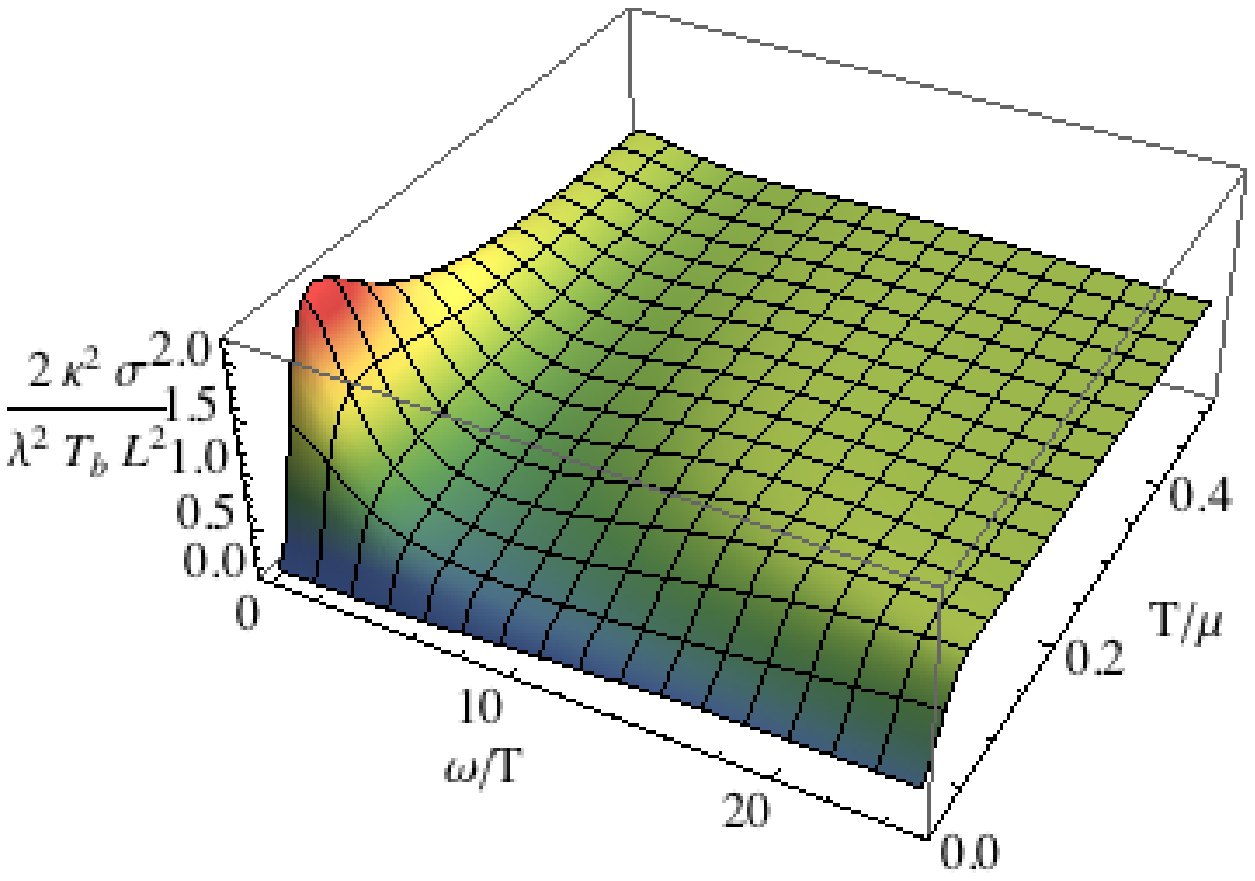}\hspace{10pt}
\includegraphics[scale=0.45]{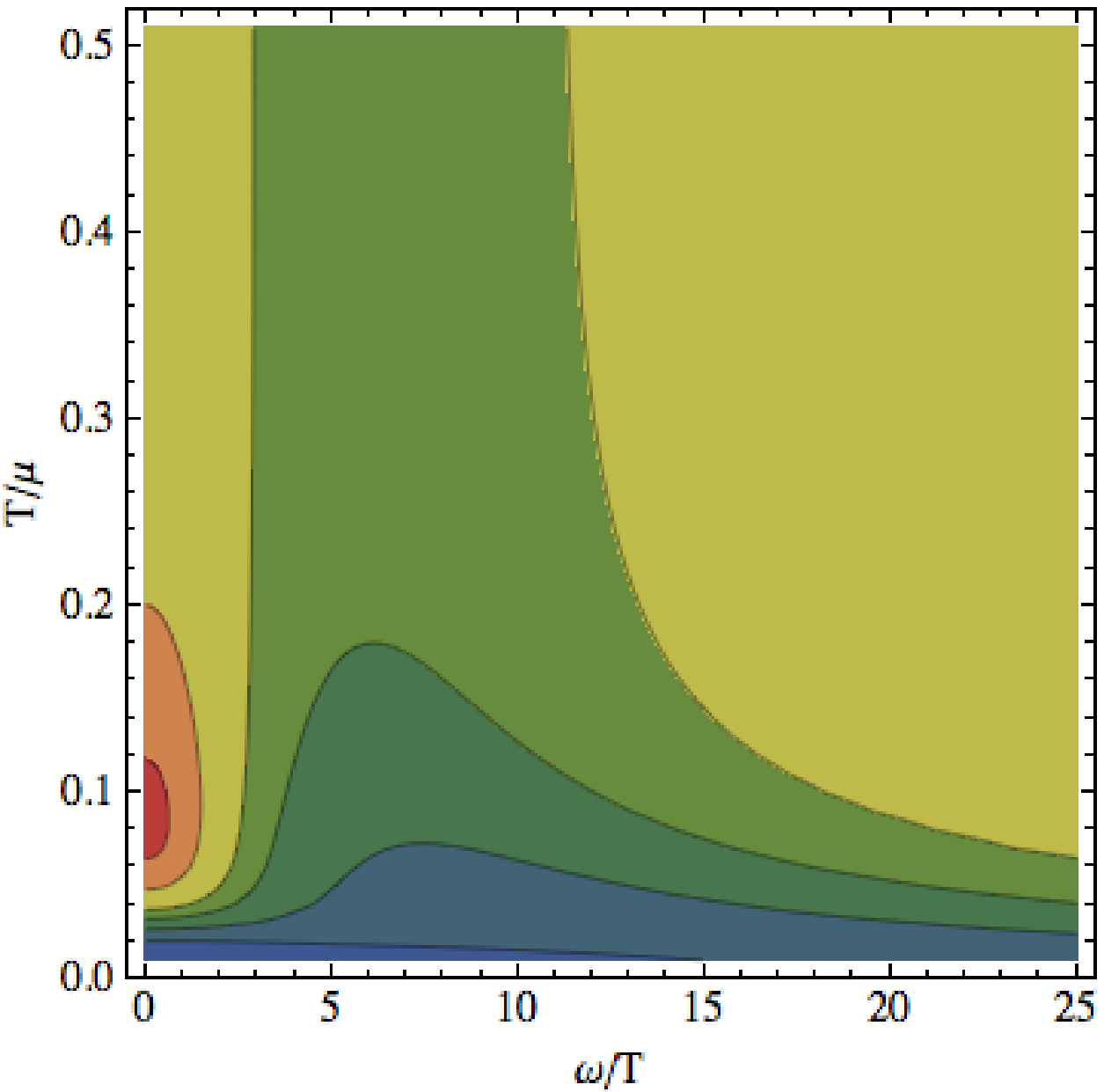}
\caption{Frequency dependence of the real part of the conductivity for $T_b \, L^2=0.1$. Both graphs are different versions of the same plot. The contours on the right-hand side graph correspond to the values of $\tfrac{2\kappa^2}{\lambda^2 T_b\, L^2}\sigma=0.25$, $0.5$, $0.75$, $1$, $1.5$ and $2$ from blue (bottom-left corner) to red (lobe  at $\tfrac{\omega}{T}=0$, $\tfrac{T}{\mu}\approx 0.09$).}\label{fig.UVfreq}
\end{center}
\end{figure}
Figure \ref{fig.UVfreq} shows a frequency dependent plot of the conductivity as a function of  $\tfrac{T}{\mu}$, at fixed $T_b \, L^2=10^{-1}$.
In the cases with $\sigma_{DC}>1$ there is a dip at medium values of the frequency.
There is a second maximum (not visible in the graph), and the  conductivity plotted in figure \ref{fig.UVfreq}  approaches $1$ from above at large values of $\tfrac{\omega}{T}$.
The dip is fundamental for the satisfaction of the sum rule
\begin{equation}\label{eq.sumrule}
\int_{0}^\infty \mathrm{Re}\left[\sigma(\omega)\right] \, d\omega = \textrm{constant} \ ,
\end{equation}
independently of $\tfrac{T}{\mu}$.
Actually, for \eqref{eq.sumrule} to be satisfied, it is also fundamental to take into consideration the delta function contribution \eqref{eq.impolehydro}.

At low frequencies  and sufficiently large values of $\tfrac{T}{\mu}$  there is a  low energy resonance  in the system which does not survive the zero temperature limit. 
This Drude-like peak is always visible in probe calculations \cite{Hartnoll:2009sz}, but backreaction effects make it disappear at low values of the temperature.

\subsubsection{Comments on the DC conductivity of the scaling solution}

In the IR scaling solutions the role of the gauge field is not to source a chemical potential in the field theory but to support the geometry \cite{Taylor:2008tg}.
 This implies that the only available scale in the game is the temperature. 
 The conductivity is proportional to $T_b$, and the power dependence on the temperature is given by dimensional analysis. 
 For example, for the value of the DC conductivity at the horizon 
\begin{equation}
\sigma_0(r_h) = \frac{\lambda^2 T_b}{2\kappa^2} \sqrt{1+\rho^2} \left(\frac{z+(d-1)(1-\gamma)}{4\pi}\right)^{\frac{d+1+\gamma(1-d)}{z}} T^{-\frac{d+1+\gamma(1-d)}{z}} \ ,
\end{equation}
and the running of this magnitude to the ultraviolet can change only  the prefactor, but not the temperature dependence.

One important example of scaling solutions is given by the $\gamma=\frac{d+1-z}{d-1}$ line (from the null energy condition  only $z\geq 2$ along this line is physical).
In this case the DC conductivity scales inversely with the temperature. 
Then the specific heat scales with the temperature as $C_V \sim T^{(z-2)/z}$ which can only be a linear dependence in the $z\to\infty$ limit. 
In this limit the geometry approaches \eqref{eq.firstapprox}, provided the ratio $\tfrac{\gamma}{z}$ is fixed.
For this geometry the DC conductivity at the horizon is
\begin{equation}
\sigma_0(r_h) = \frac{\lambda^2 T_b}{2\kappa^2} \sqrt{1+ z^4 \rho^2} \left(\frac{z\, \alpha}{4\pi}\right)^{\alpha-1} T^{1-\alpha} \ .
\end{equation}
The specific heat scales with the temperature as $C_V\sim T^{\alpha-1}$.
The value  $\alpha=2$ is appropriate for the description of non-Fermi liquids from the temperature behaviour of the conductivity and the specific heat.

\subsection{Scalar channel}

Only the UV-complete solution is studied in the rest of the section.
The perturbations  in the scalar channel comprise the fluctuation of the scalar, the longitudinal electric field, and the sound fluctuations of the metric.
Actually, for the UV-complete solution the perturbation of the dilaton decouples, and will not be considered further.
The equations of motion are solved numerically following the strategy in \cite{Kaminski:2009dh}.
An ingoing-wave boundary condition at the horizon is imposed, and the two equations of motion integrated with two independent normalisations of the fields at the horizon.
The most general solution is a linear combination of the two independent solutions (each solution gives a radial profile for each of the two fields) obtained in this way, and the combination that gives direct information about the two-point correlators of the theory corresponds to a particular normalisation of the fields at the boundary.

In this section the focus is only on quasinormal modes (QNM), which are given by the values of the (complex) frequency where the boundary normalisation is not well defined: the non-normalizable mode of the fields vanishes, implying a vanishing source of the dual operator in the field theory.
The numerical recipe to find the position of the QNM is to take the two independent solutions normalised at the horizon (each solution organised as a column vector), take the determinant of the $2\times2$ matrix constructed with them and look for the frequencies at which this determinant vanishes at the boundary.

In the present case the equations of motion for the longitudinal electric field and the metric fluctuations are coupled, signalling an RG mixing of the operators that these field describe.
This in turn means that the QNM cannot be associated to one of the fluctuations individually: all fluctuations share the same QNM.

\vspace{15pt}

In a thermal setup, for $\tfrac{\omega}{T} \sim \tfrac{q}{T} \ll \tfrac{T}{\mu}$ hydrodynamic modes dominate the large time behaviour of the system.
The longitudinal fluctuation of the electric field gives information about the diffusion of charge in the system via a hydrodynamic pole with dispersion relation
\begin{equation} \label{eq.diffusivepole}
\omega = - i \, {\cal D} \, q^2  + {\cal O}(q^4)\ ,
\end{equation} 
with $\cal D$ the diffusion constant, as described by Fick's law $\partial_t \phi = {\cal D} \partial_x^2 \phi$.\footnote{Notice that in the vectorial channel there is also a diffusive mode with dispersion relation \eqref{eq.diffusivepole}, with a different diffusion constant. In this case the mode describes diffusion of momentum. This mode is not considered in this paper.}
The hydrodynamic mode associated to the scalar perturbations of the metric consists of the sound mode, with dispersion relation
\begin{equation}\label{eq.soundmode}
\omega = c \, q - i \, \Gamma\, q^2 + {\cal O}(q^3) \ ,
\end{equation}
where the hydrodynamic prediction for $c$ is the speed of sound \eqref{eq.speedofsound}, and for the attenuation coefficient is $\Gamma=\tfrac{\eta}{{\cal E}+P}$.
In holographic setups based on Einstein gravity with space isotropy, like the model studied here,  the shear viscosity is related to the entropy via the KSS value $\tfrac{\eta}{s}=\tfrac{1}{4\pi}$ \cite{Kovtun:2004de,Iqbal:2008by}.
This value is used below when evaluating the attenuation coefficient.

\vspace{15pt}

\begin{figure}[tb]
\begin{center}
\includegraphics[scale=0.6]{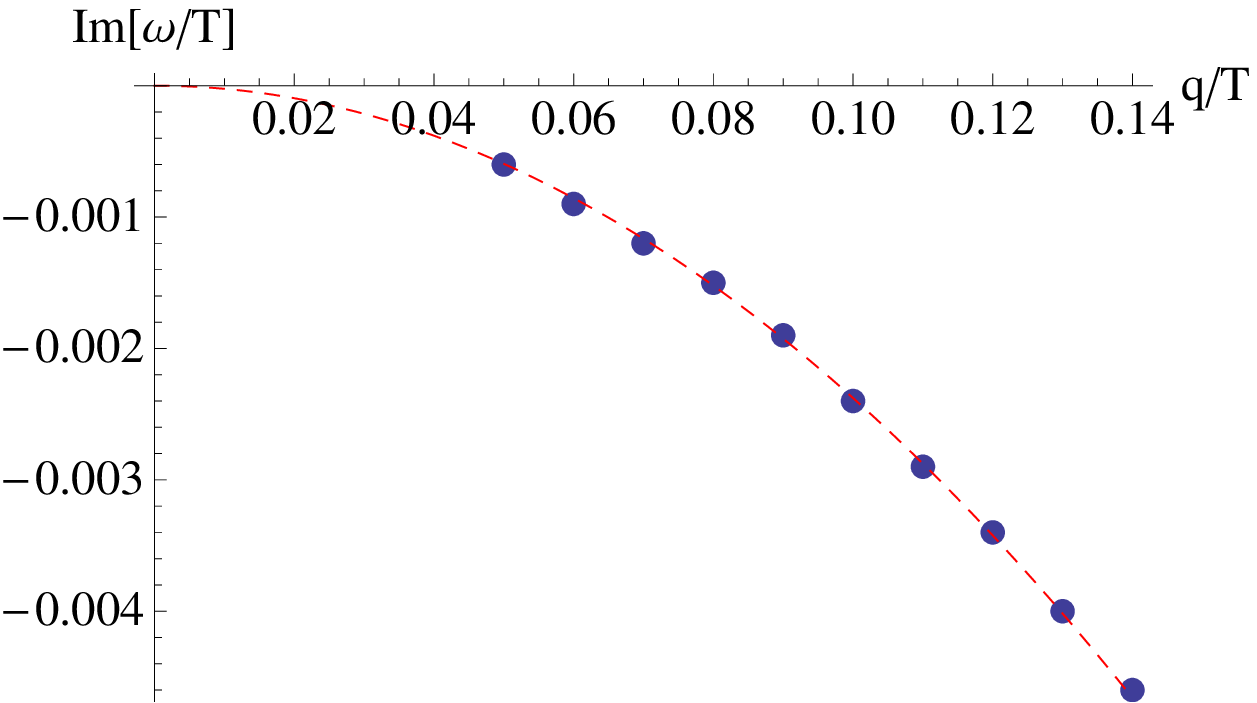}
\includegraphics[scale=0.6]{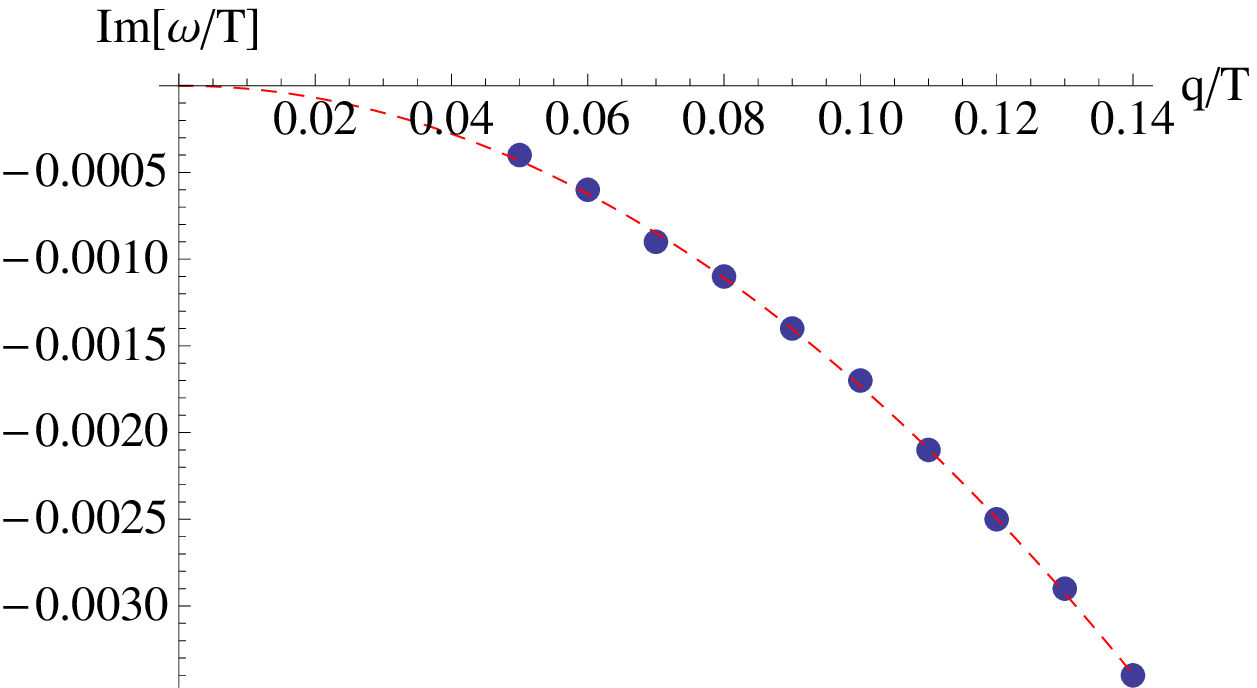}
\caption{Frequency of the charge diffusion pole for different values of the momentum for $T_b \, L^2=2$ and $\tfrac{T}{\mu}=2$ (left) and $\tfrac{T}{\mu}=0.2$ (right). The dashed line is the Einstein relation prediction for the diffusion constant.}\label{fig.Tb2diffusion}
\end{center}
\end{figure}

\begin{figure}[tb]
\begin{center}
\includegraphics[scale=0.4]{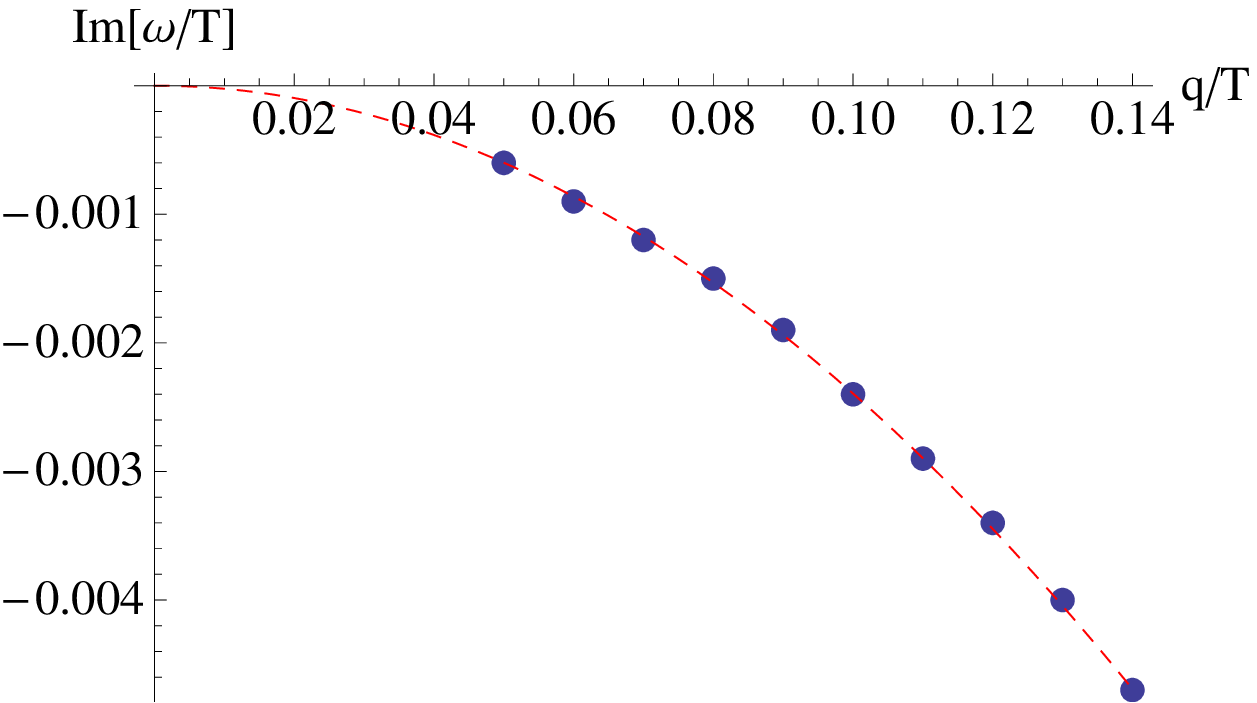}
\includegraphics[scale=0.4]{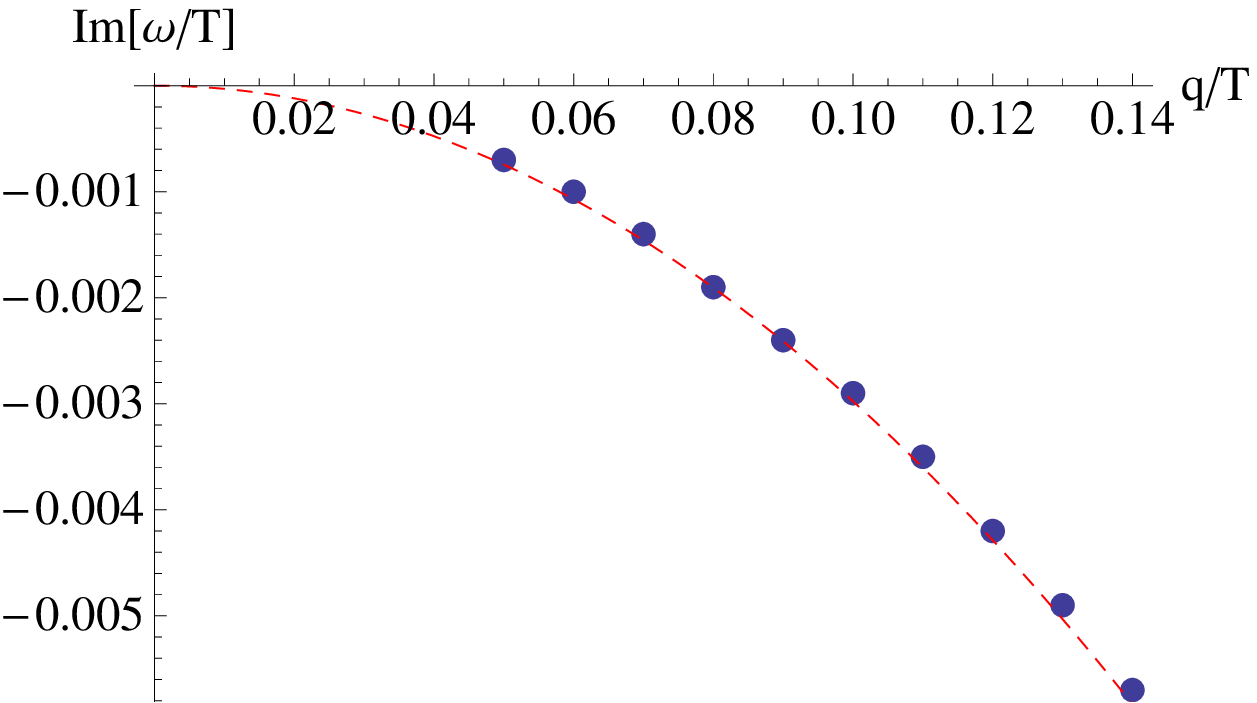}
\includegraphics[scale=0.4]{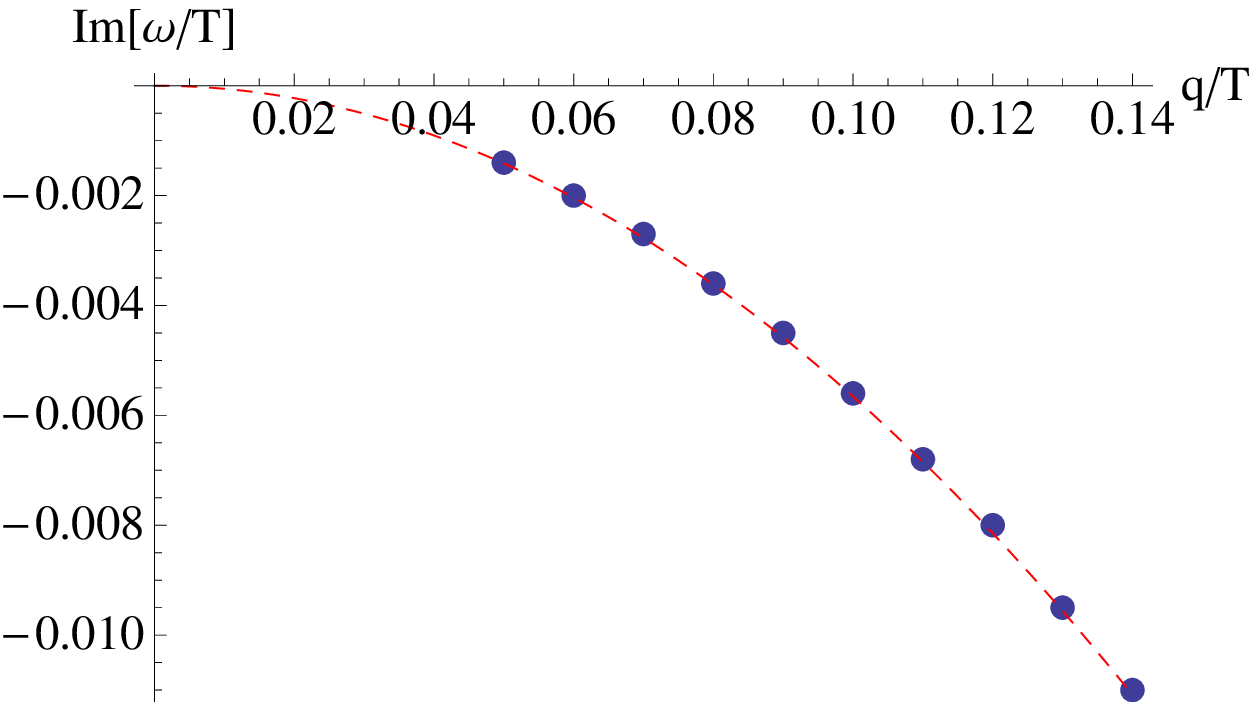}
\caption{Frequency of the charge diffusion pole for different values of the momentum for $T_b \, L^2=0.01$ and $\tfrac{T}{\mu}=2$ (left), $\tfrac{T}{\mu}=0.2$ (center) and $\tfrac{T}{\mu}=0.02$ (right). The dashed line is the Einstein relation prediction for the diffusion constant.}\label{fig.Tb001diffusion}
\end{center}
\end{figure}

An example of the diffusive dispersion relation for a relatively large value of $T_b\, L^2$ is found in figure \ref{fig.Tb2diffusion}.
There is a quadratic dependence of the frequency with the momentum of the fluctuation.
Einstein's relation  (the dashed line in the figures for the diffusive dispersion relation) gives a good estimation of the diffusion constant, relating the thermodynamic quantities obtained in section \ref{sec.UV}, the DC conductivity \eqref{eq.myILeq}, and the diffusion charge constant \cite{Kovtun:2012rj}
\begin{equation} \label{eq.ERel}
{\cal D} = V_{d-1}  \frac{\sigma_{DC} \left({\cal E + P} \right)^2 }{  \left({\cal E + P} \right)^2 \chi - 2 \, Q \, \left({\cal E + P} \right)  \left( \tfrac{\partial {\cal E}}{\partial \mu} \right)_{T} + Q^2 T \left( \tfrac{\partial {\cal E}}{\partial T} \right)_{\mu/T}} \ ,
\end{equation}
where we have multiplied by the volume of $\mathbb{R}^{d-1}$ to cancel the factor appearing in the susceptibilities in the denominator.
When the value of $T_b\, L^2$ is lowered, Einstein's relation still holds, even for low temperatures, as seen in figure \ref{fig.Tb001diffusion}.

Relation \eqref{eq.ERel} in probe calculations holds with $Q$ the value of the background spacetime, not the value of the charge in the probe \cite{Mas:2008qs}.

\vspace{15pt}

\begin{figure}[tb]
\begin{center}
\includegraphics[scale=0.5]{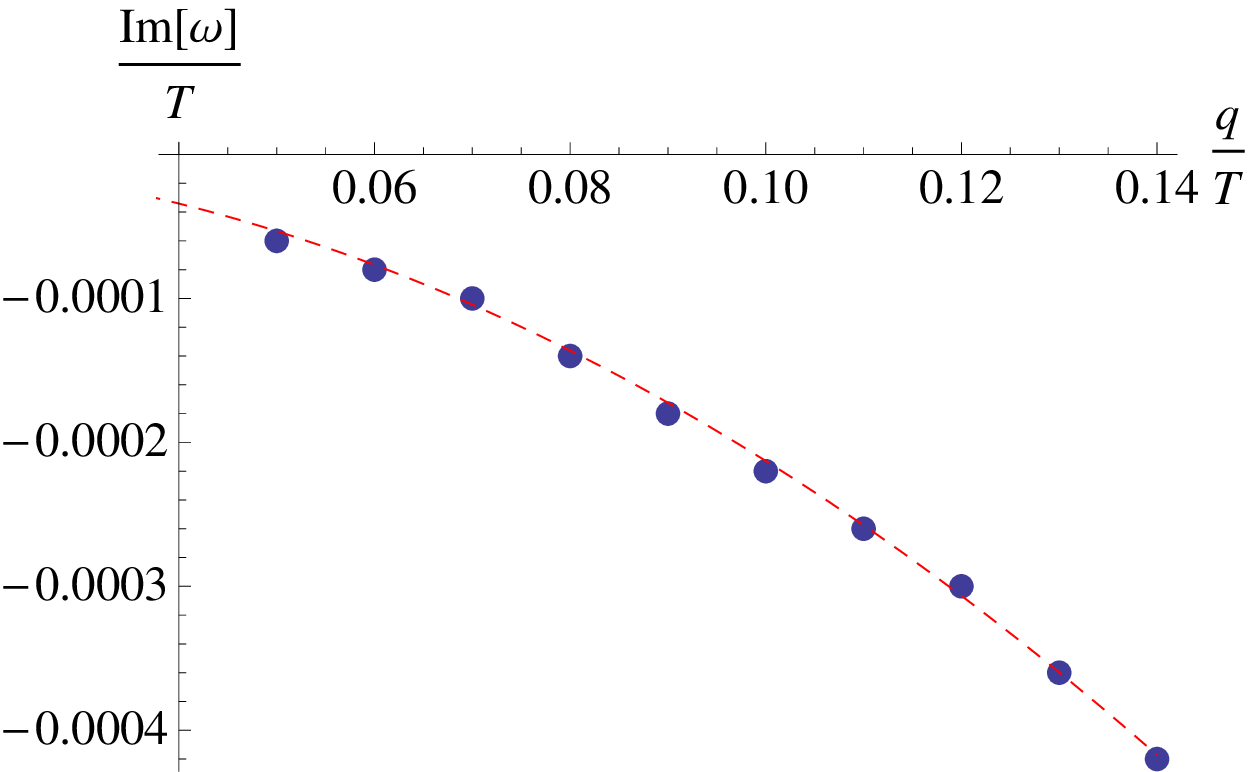}
\includegraphics[scale=0.5]{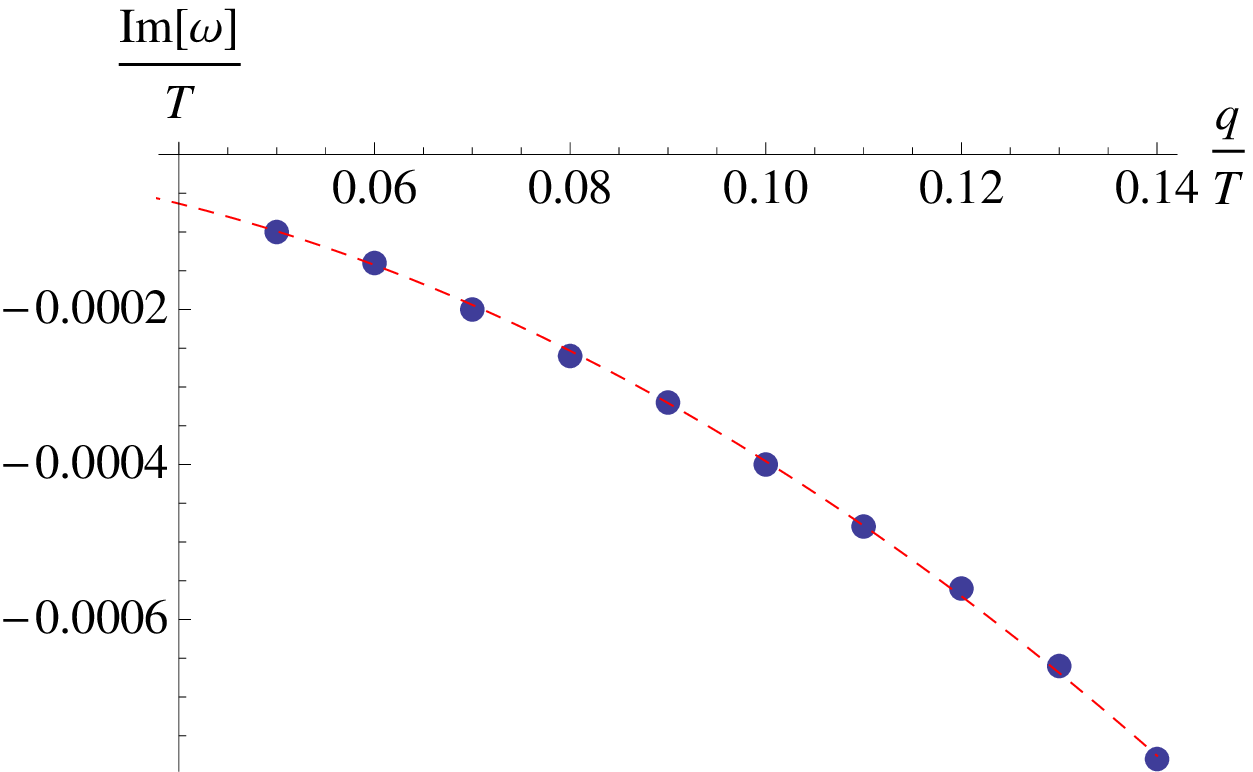}
\includegraphics[scale=0.5]{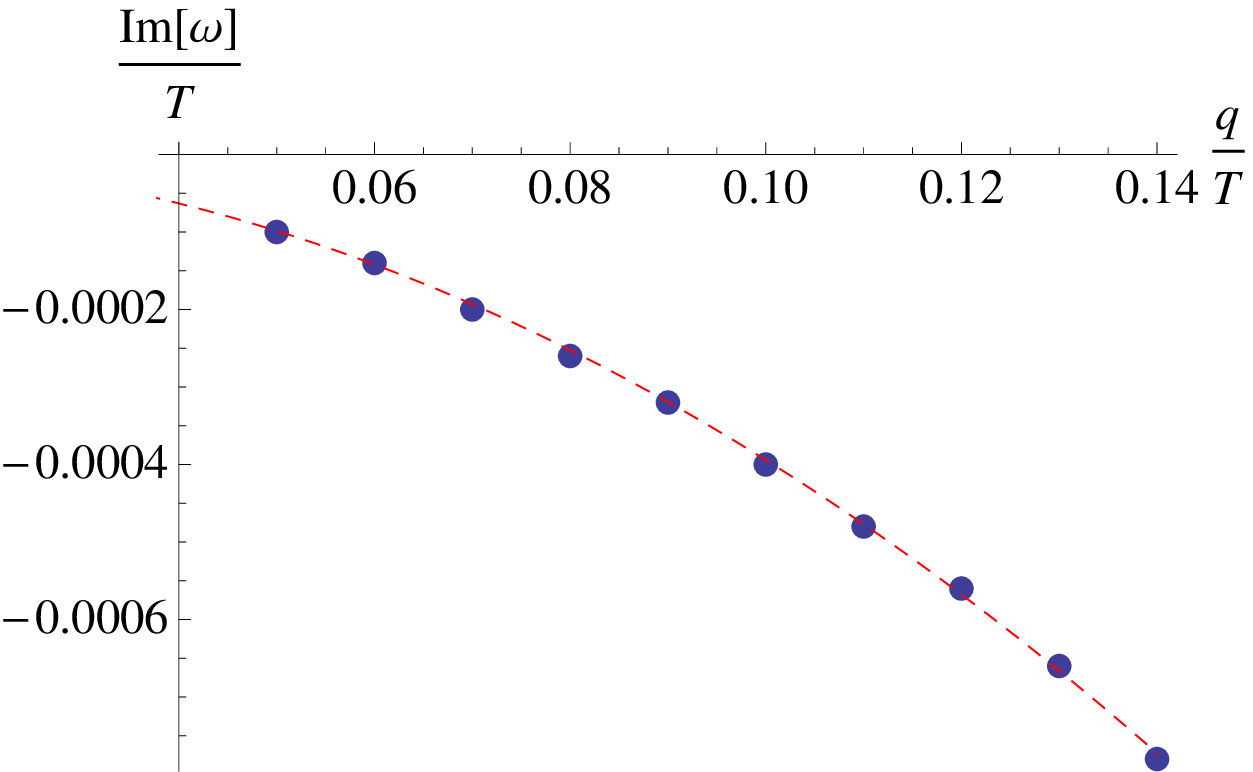}
\includegraphics[scale=0.5]{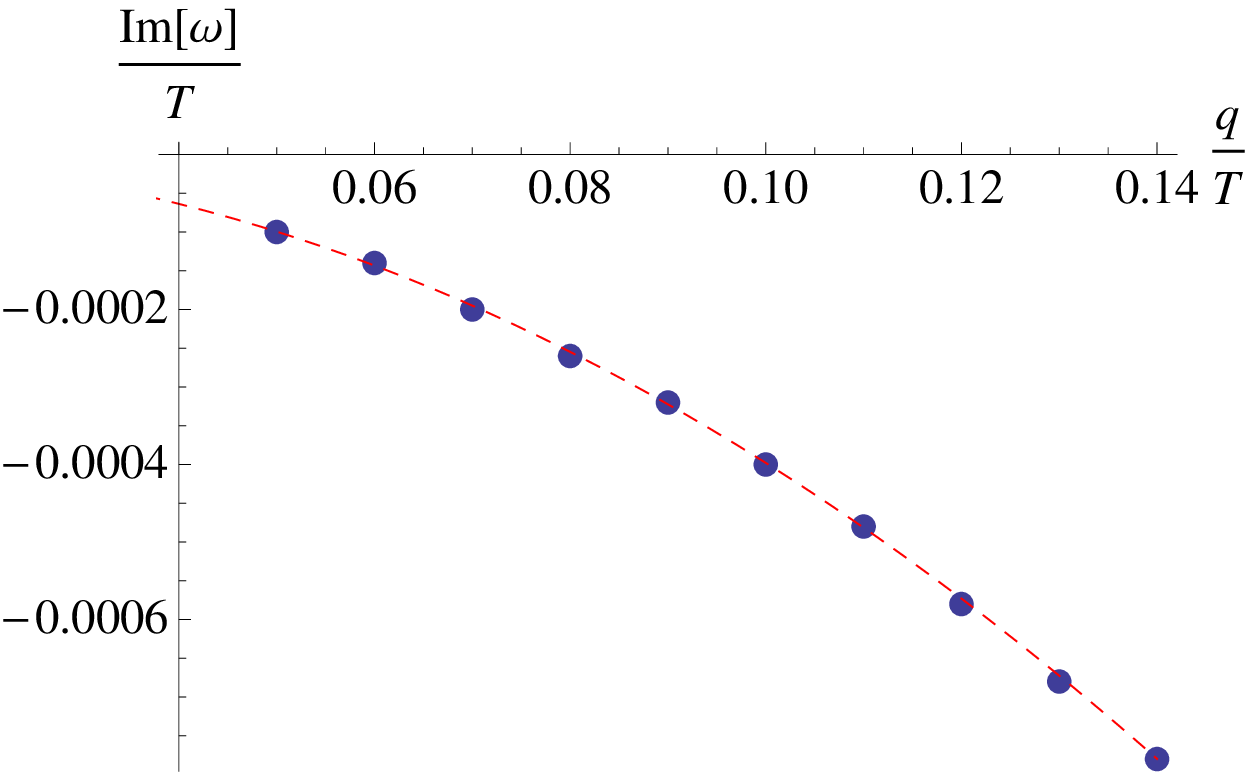}
\caption{Imaginary part of the sound pole for different values of the momentum. The upper row corresponds to $\tfrac{T}{\mu}=0.2$, and the lower row to $\tfrac{T}{\mu}=2$. The left column is for $T_b \, L^2=2$ and the right column for $T_b \, L^2=0.01$. The dashed lines are the attenuation factors predicted by hydrodynamics. }\label{fig.attenuationshydro}
\end{center}
\end{figure}

The sound mode has a dispersion relation linear in the momentum for the real part and quadratic for the imaginary part. 
The real part is accurately described by the speed of sound $c=c_s=\tfrac{1}{\sqrt{2}}$ (no graphs are reported).
As for the attenuation coefficient $\Gamma$, in figure \ref{fig.attenuationshydro} different graphs are given where the dependence of $\mathrm{Im}[\omega]$ with  the momentum, in the hydrodynamic limit $\tfrac{\omega}{T} \sim \tfrac{q}{T} \ll \tfrac{T}{\mu}$, is shown.
In these figures the dashed lines correspond to the value of $\Gamma$  given by hydrodynamics, which in the present context reads
\begin{equation}\label{eq.attenuationT}
\Gamma \, T= \frac{\eta}{{\cal E}+P} \frac{d-2}{d-1}\,T = \frac{d-2}{d-1} \frac{r_+^d}{4\, \pi\, m} \left( 1 + T_b \, L^2 \frac{1 - \sqrt{1 + \varrho^2}}{d(d-1)} \right) \ ,
\end{equation}
where $m$ is given in \eqref{eq.mcoeff} and $\varrho$ in equation \eqref{eq.varrho}. 
Notice that $m \propto r_+^d$, such that the attenuation coefficient times the temperature is not explicitly dependent on the size of the horizon.
The temperature dependence enters via the quotient $\tfrac{T}{\mu}$  controlled by $\varrho$.
Except for the upper-left plot in figure \ref{fig.attenuationshydro}, all attenuation factors are very similar in these graphs.

\vspace{15pt}

Holographic charged systems at zero temperature have a collective excitation, the so-called holographic zero sound, which can be found as a QNM in the retarded correlator \cite{Karch:2008fa}.
Later, it was found \cite{Davison:2011ek} that this excitation survives at low temperatures, with a dispersion relation still dominated by the extremal result.
This is the collisionless quantum regime, $1\ll \tfrac{\omega}{T} \ll \tfrac{\mu}{T}$.

\begin{figure}[tb]
\begin{center}
\includegraphics[scale=0.6]{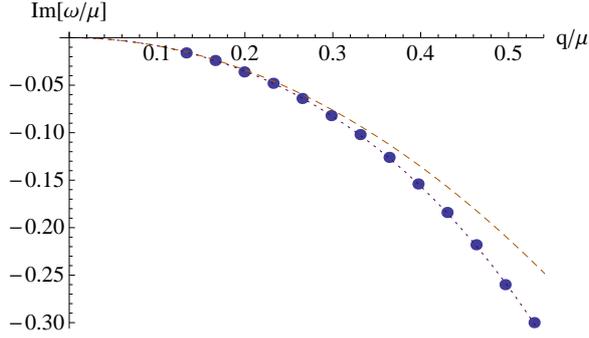}
\caption{Frequency of the charge diffusion pole for different values of the momentum for $T_b \, L^2=1$ and $\tfrac{T}{\mu}=0.2$. The dashed line is the quadratic term of the  numeric fit to the data  $\tfrac{\textrm{Im}[\omega]}{\mu} = -i \, 0.8482 \frac{q^2}{\mu^2} - i \, 0.8103 \frac{q^4}{\mu^4}$. The quartic fit is given by the dotted line that passes over the data points.}\label{fig.zerodiffusion}
\end{center}
\end{figure}

As in the thermal-dominated regime, for low temperatures there are two poles: one diffusive pole with dispersion relation \eqref{eq.diffusivepole} and one sound mode with dispersion relation \eqref{eq.soundmode}.
These modes were studied in the AdS$_4$ Reissner-Nordstr\"om case in \cite{Edalati:2010pn} at zero temperature and in \cite{Davison:2011uk} at finite $\tfrac{T}{\mu}$.
In the linear electrodynamics case,  at zero temperature  there is also a branch cut that hides the existence of the  diffusion mode, and hydrodynamics is not enough to describe the full theory \cite{Edalati:2010pn,Davison:2013bxa}.\footnote{I would like to thank Richard A. Davison for  comments on this.}
The hydrodynamic mode is still described by Einstein's relation \eqref{eq.ERel}: both the numerator and denominator in \eqref{eq.ERel} behave as $T^2$ at low temperatures, giving a finite diffusion constant.
In figure \ref{fig.zerodiffusion} the position of the diffusive mode in the imaginary frequency axis is plotted for a low value of the temperature, but with the momentum given in units of the chemical potential for convenience, such that the dominant scale is the chemical potential.
Notice how in figure \ref{fig.zerodiffusion} the quartic term in the dispersion relation, which was not written in \eqref{eq.diffusivepole}, is necessary to obtain a good description for values greater than $\tfrac{q}{\mu}\sim 0.2$. 

\begin{figure}[tb]
\begin{center}
\includegraphics[scale=0.6]{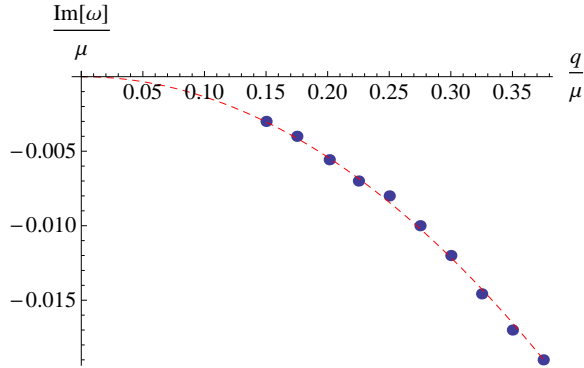}
\caption{Imaginary part of the sound pole for different values of the momentum with $T_b \, L^2=1$ and $\tfrac{T}{\mu}=0.2$. The dashed line is the attenuation factor predicted by hydrodynamics.}\label{fig.zerosound}
\end{center}
\end{figure}

For the sound mode, the real part is still accurately given by the speed of sound, as was always the case in the AdS-RN case \cite{Davison:2011ek}.
The attenuation factor is also determined by hydrodynamics, as shown in figure \ref{fig.zerosound}.
Recall that the result in equation \eqref{eq.attenuationT} is multiplied by the temperature, since in the corresponding section the dimensionless frequency and momentum are measured with respect to it: $\tfrac{\omega}{T}$ and $\tfrac{q}{T}$.
In figure \ref{fig.zerosound}, however, the dimensionless quantities that are shown are changed to $\tfrac{\omega}{\mu}$ and $\tfrac{q}{\mu}$ for convenience, and the dashed line in the figure is given by the expression in \eqref{eq.attenuationT} multiplied by  $\tfrac{\mu}{T}$.

\section{Conclusions}\label{sec.conclusions}

In this paper we have considered the system given by the Einstein-Hilbert action coupled to a scalar and a DBI-type action. 
In particular, we have considered  solutions describing a charged, asymptotically-AdS black hole with trivial scalar, and a scaling solution with an asymptotically hyperscaling-violating, Lifshitz metric, and running dilaton.
These solutions are not new \cite{Pal:2012zn}, however the thermodynamic analysis, together with the holographic renormalization of the UV-complete solution and the background subtraction in the scaling solution appear here for the first time.
Focusing on the UV-complete solution of section \ref{sec.UV}, we have checked that in many senses this is a generalisation of the AdS-Reissner-Nordstr\"om solution to nonlinear electromagnetism.
In particular, the zero temperature entropy (density) is finite, the stress-energy tensor is traceless and the specific heats and susceptibility are non-negative.

Focusing on $d=3$, we have studied different transport properties of the  field theory dual to the UV-complete solution.
The system possesses hydrodynamic modes that drive the behaviour of the system at larger times.
In particular there are sound modes (whose dispersion relation is given by hydrodynamics) diffusion of charge modes  and diffusion of momentum modes.

The retarded current-current correlator gives the electric conductivity.
To calculate it we have adapted the prescription in \cite{Iqbal:2008by} to include a mass term in the equation of motion of the fluctuation dual to the current operator.
This allowed  to determine the strength of the zero-frequency pole in the imaginary part of the conductivity, which is related to a delta function in the real part, and the value of the $\omega\to0$ (DC) limit of the conductivity by integration of first order differential equations.
At large values of the temperature (compared to the chemical potential) the DC conductivity of the field theory is well approximated by the value of the bulk conductivity at the horizon, i.e., the running with the bulk radius is highly suppressed.
At lower temperatures the RG flow is important, and gives a DC conductivity that vanishes with the temperature as $T^2$ for $d=3$.
The behaviour of the conductivity shows explicitly that backreaction of the charged brane onto the geometry is important at low values of the temperature.

\vspace{15pt}

There are several possible continuations to the work presented here.

 A study in the strict $T=0$ case, similar to the one performed in \cite{Edalati:2010pn,Edalati:2009bi,Edalati:2010hk}, would be  interesting to study the effects of non-linearities in the extremal case, and in particular in the existence of a branch cut along the imaginary frequency axis in the scalar channel.

We have not considered the scalar fluctuations of the Lifshitz scaling solution, although the equations of motions presented here are ready for such a study. 
It would be interesting to calculate these, and study potential flows that have this geometry in the IR of the theory, but an asymptotically-AdS one in the UV.

 From the point of view of the model, one obvious generalisation is to allow a finite codimension of the brane, in a similar way to the actions used in \cite{Chang:2013mca,Jensen:2013lxa}.

A different generalisation is to consider an asymptotically AdS solution with no trivial dilaton. One such possibility is to take the models of IHQCD \cite{Jarvinen:2011qe,Alho:2012mh} and adapt them to include a chemical potential in the way showed in this work. This would mimic the presence of a finite baryon density in strongly coupled QCD.

\section*{Acknowledgements}

I would like to thank Daniel Are\'an for useful discussions and comments on a previous version of the draft, and specially
Richard A. Davison for pointing out references \cite{Hartnoll:2007ih,Hartnoll:2007ip,Kovtun:2012rj} and his comments on the first version of this draft.

I would like to thank also Ant\'on Faedo and David Mateos for helpful discussions. 
I am supported by MEC FPA2010-20807-C02-02, by ERC StG HoloLHC - 306605 and by the Juan de la Cierva program of
the Spanish Ministry of Economy.

\appendix

\section{Scalar fluctuations' equations of motion}  \label{app.equations}

In this appendix  the  equations of motion for the fluctuations in the scalar sector under the $SO(d-2)$ little group are given, before taking  the diffeo+gauge invariant combinations.

These are four second-order equations coming from the Einstein equations:
\begin{align}\label{eq.app.scal1}
0  = & H_{00}'' + \partial_r \log \left[ \sqrt{\frac{g_{tt}^2g_{xx}^{d-1}}{g_{rr}}} \right] H_{00}'  + \partial_r \log \left[ \sqrt{g_{tt}} \right]  \left( H_{11}'  + (d-2)  H_{ii}' \right) \nonumber  \\
& - \frac{T_b\,  \rho\, g_{rr} \left( (d-3)  \rho^2 + 2(d-2) Z_1^{2}Z_2^{d-1} g_{xx}^{d-1} \right)}{ Z_1^{2}Z_2^{d-1} (d-1)  g_{xx}^{\frac{3}{2}(d-1)} }  \left( \frac{ \lambda\, \alpha_0' }{ \sqrt{g_{rr} g_{tt} } } - \frac{Z_2\, \rho}{2\sqrt{\rho^2 + Z_1^2 Z_2^{d-1} g_{xx}^{d-1} } } H_{00} \right) \nonumber \\
& + g_{rr} \left(-  \frac{q^2}{g_{xx}}  H_{00} + \frac{2\, q\, \omega }{g_{tt}} H_{01} +   \frac{ \omega^2 }{g_{tt}} \left( H_{11} + (d-2) H_{ii} \right) \right) \nonumber \\
& + \frac{ g_{rr}}{d-1} \Bigg(  2\, \partial_\phi V + \frac{Z_1^3 Z_2^{-(d-1)} T_b}{2 g_{xx}^{\frac{3}{2}(d-1)} \sqrt{  \rho^2 + Z_1^2 Z_2^{d-1} g_{xx}^{d-1} } }  \Big( 2(d-3) e^{(5\alpha+\beta)\phi}\beta\, \rho^4 \nonumber \\
& \qquad\qquad\qquad+ e^{(3\alpha+d\beta)\phi} (2 (d-3) \alpha-(d-5)(d-1)\beta )\rho^2\, g_{xx}^{d-1} \nonumber \\
& \qquad\qquad\qquad - 2 e^{(\alpha + (2d-1)\beta)\phi} (2\alpha - (d+1) \beta) g_{xx}^{2(d-1)}   \Big)  \Bigg) \varphi \ ,
\end{align}
and
\begin{align}
0 = & H_{11}'' + \partial_r \log \left[ \sqrt{  \frac{ g_{tt}\,g_{xx}^d }{ g_{rr} } }   \right] H_{11}' + \partial_r \log \left[ \sqrt{g_{xx}} \right] \left( H_{00}'+(d-2)H_{ii}' \right)  \nonumber \\
& + 2 g_{rr} \, \frac{T_b\, \rho\, \left(  \rho^2 + Z_1^{2}Z_2^{d-1} g_{xx}^{d-1} \right)}{Z_1^{2}Z_2^{d-1} (d-1) \, g_{xx}^{\frac{3}{2}(d-1) }}   \left( \frac{ \lambda\, \alpha_0' }{ \sqrt{g_{rr} g_{tt} } } - \frac{Z_2\, \rho}{2\sqrt{\rho^2 + Z_1^2 Z_2^{d-1} g_{xx}^{d-1} } } H_{00} \right) \nonumber \\
& + g_{rr} \left(  \frac{2 \, q \, \omega}{g_{tt}} H_{01} + \frac{\omega^2 }{g_{tt}} H_{11} -   \frac{q^2}{g_{xx}}  \left( H_{00} + (d-2) H_{ii} \right) \right) \nonumber \\
& +\frac{g_{rr}}{d-1} \Bigg(2\, \partial_\phi V - \frac{Z_2^{-(d-2)} T_b\, \sqrt{\rho^2 + Z_1^2Z_2^{(d-1)} g_{xx}^{d-1} }  }{ g_{xx}^{\frac{3}{2}(d-1)} } \Big( 2e^{2\alpha\phi} \beta\, \rho^2 \nonumber \\
& \qquad\qquad\qquad + e^{(d-1)\beta\phi} (2\alpha - (d+1)\beta) g_{xx}^{d-1}  \Big) \Bigg) \varphi \ ,
\end{align}
and
\begin{align}
0 = & H_{ii}'' + \partial_r \log \left[ \sqrt{ \frac{ g_{tt}\, g_{xx}^{2d-3} }{ g_{rr} } } \right] H_{ii}' + \partial_r \log \left[ \sqrt{ g_{xx}} \right] \left( H_{00}' + H_{11}' \right) + g_{rr} \left(  \frac{\omega^2}{g_{tt}} - \frac{q^2}{g_{xx}} \right) H_{ii} \nonumber \\
& + 2\frac{T_b \, \rho\, g_{rr} \left(  \rho^2 + Z_1^{2} Z_2^{d-1} g_{xx}^{d-1} \right) }{Z_1^{2} Z_2^{d-1} (d-1)\, g_{xx}^{\frac{3}{2}(d-1)} }  \left( \frac{ \lambda\, \alpha_0' }{ \sqrt{g_{rr} g_{tt} } } - \frac{Z_2\, \rho}{2\sqrt{\rho^2 + Z_1^2 Z_2^{d-1} g_{xx}^{d-1} } } H_{00} \right)   \nonumber \\
& + \frac{g_{rr}}{d-1}\Bigg( 2\, \partial_\phi V - \frac{Z_2^{-(d-2)} T_b\,  \sqrt{ \rho^2 + Z_1^2 Z_2^{(d-1) } g_{xx}^{d-1} }  }{  g_{xx}^{\frac{3}{2}(d-1)} } \Big( 2e^{2\alpha\phi} \beta\, \rho^2 \nonumber \\
& \qquad\qquad\qquad+ e^{(d-1)\beta\phi} (2\alpha - (d+1)\beta) g_{xx}^{d-1}  \Big) \Bigg) \varphi \ ,
\end{align}
and
\begin{equation}\label{eq.app.scal2}
0= H_{01}'' + \partial_r \log \left[  \sqrt{ \frac{ g_{xx}^{d+1} }{ g_{tt} g_{rr} } } \right] H_{01}' + T_b\, \lambda\, \rho\, \sqrt{\frac{ g_{rr} g_{tt} }{ g_{xx}^{d+1} }} \alpha_1' + (d-2)\,q\,\omega \frac{g_{rr}}{g_{xx}} H_{ii} \ .
\end{equation}

Two second-order equations of motion coming from the vector field fluctuations:
\begin{align}\label{eq.app.scal3}
0= & \alpha_0'' + \partial_r \log \left[ \sqrt{ \frac{ \left( \rho^2  +  Z_1^{2} Z_2^{d-1} g_{xx}^{d-1} \right)^3 } { g_{tt}g_{rr} g_{xx}^{2(d-1)} } } Z_1^{-2} Z_2^{-d} \right] \alpha_0' -q\, \frac{Z_1^{2} Z_2^{d-1} \,g_{rr} g_{xx}^{d-2}}{\rho^2 +Z_1^{2} Z_2^{d-1} g_{xx}^{d-1}} \left( \omega\, \alpha_1+ q \, \alpha_0 \right) \nonumber \\
& + \frac{Z_1^{2}Z_2\, \rho}{2\lambda} \sqrt{ \frac{g_{rr} g_{tt} }{ (  \rho^2 + Z_1^{2} Z_2^{d-1} g_{xx}^{d-1} )^3 } } \Bigg(  - (e^{2\alpha\phi}\rho^2 + e^{{d-1}\beta\phi} g_{xx}^{d-1}) H_{00}' \nonumber\\
& \qquad   + e^{(d-1)\beta\phi} g_{xx}^{d-1} (H_{11}'+(d-2)H_{ii}' )   - (2e^{2\alpha\phi} \beta\, \rho^2 + e^{(d-1)\beta\phi} (2\alpha-(d-3)\beta) g_{xx}^{d-1} ) \varphi' \Bigg) \nonumber \\
& + Z_2 \rho^3 \sqrt{g_{rr}g_{tt}} \frac{(d-1)g_{xx}' - (2\alpha-(d-1)\beta) g_{xx} \phi'}{\lambda g_{xx}(\rho^2 + Z_1^2 Z_2^{d-1} g_{xx}^{d-1})^\frac{3}{2} } \left( \frac{H_{00}}{2} + \beta \, \varphi  \right)  \ ,
\end{align}
and
\begin{align}\label{eq.app.scal4}
0 = & \alpha_1''+\partial_r \log \left[ \sqrt{ \frac{ g_{tt} }{ g_{rr}g_{xx}^2 } } \sqrt{  \rho^2  +Z_1^{2} Z_2^{d-1}  g_{xx}^{d-1} }  Z_2 \right] \alpha_1' + \omega \frac{g_{rr}}{g_{tt}} (\omega \alpha_1+q \alpha_0) \nonumber \\
& + \frac{ Z_2 \rho\, g_{xx} }{\lambda} \sqrt{ \frac{ g_{rr} }{ g_{tt} (\rho^2  +Z_1^{2}  Z_2^{d-1}g_{xx}^{d-1} ) } } H_{01}' \ .
\end{align}

The second-order equation of motion for the scalar perturbation:
\begin{align}\label{eq.app.scal5}
0 = & \varphi'' + \partial_r \log \left [ \sqrt{  \frac{ g_{tt} g_{xx}^{d-1} }{ g_{rr} }  } \right] \varphi' + \frac{\phi'}{2} \left( H_{00}' +H_{11}' + (d-2) H_{ii}' \right) \nonumber \\
& - T_b \lambda\,\rho \, g_{rr}  \frac{ (2\alpha - (d-3) \beta) g_{xx}^{d-1} + 2 \beta\, e^{(2\alpha-(d-1)\beta)\phi}\rho^2 }{ 2\sqrt{g_{rr}g_{tt}} g_{xx}^{\frac{3}{2}(d-1)} } \, \alpha_0' \nonumber \\
& + T_b\, \rho^2g_{rr}\, e^{(\alpha-(d-2)\beta)\phi} \frac{ 2 e^{2\alpha\phi} \beta\, \rho^2 + e^{(d-1)\beta\phi} (2\alpha - (d-3) \beta) g_{xx}^{d-1} }{ 4 \sqrt{ e^{2\alpha\phi}\rho^2 + e^{(d-1)\beta\phi} g_{xx}^{d-1} } g_{xx}^{\frac{3}{2}(d-1)} } \, H_{00} \nonumber \\
& +g_{rr} \Bigg(  \frac{ \omega^2}{g_{tt}} - \frac{q^2}{g_{xx}} - \partial_\phi^2 V  + \frac{T_b \, e^{-(\alpha+(d-1)\beta)\phi}}{ 4 \sqrt{ e^{2\alpha\phi}\rho^2 + e^{(d-1)\beta\phi} g_{xx}^{d-1} } g_{xx}^{\frac{3}{2}(d-1)} } \Big( 4 e^{(4\alpha+\beta)\phi}\beta^2\rho^4 \nonumber \\
& \qquad  \qquad +  4 e^{(2\alpha+d \beta)\phi} (2\alpha-(d-1)\beta)\beta\, \rho^2 g_{xx}^{d-1} - e^{(2d-1)\beta\phi} (2\alpha-(d+1)\beta)^2 g_{xx}^{2(d-1)}  \Big) \Bigg) \varphi\ .
\end{align}

A non-independent first-order constraint from the radial component of the Maxwell equations:
\begin{align}\label{eq.app.scal6}
0 = & \frac{2  Z_2 q\,\lambda}{\rho\, g_{xx}} \sqrt{ \frac{g_{tt}}{g_{rr}} } \sqrt{ \rho^2  + Z_1^{2} Z_2^{d-1} g_{xx}^{d-1} } \alpha_1' + \frac{ 2 Z_1^{-2} Z_2^{-d} \lambda\, \omega (\rho^2  + Z_1^{2}  Z_2^{d-1} g_{xx}^{d-1})^{3/2} }{ \rho \sqrt{g_{rr} g_{tt}}\, g_{xx}^{d-1} } \alpha_0' \nonumber \\
& +  2 q\, H_{01} +\omega \left( H_{11} + (d-2) H_{ii}   \right) - \omega \frac{ \rho^2 + Z_1^{2} Z_2^{d-1} g_{xx}^{d-1} }{ Z_1^{2} Z_2^{d-1} g_{xx}^{d-1} }  H_{00}\nonumber \\
& +\omega  \left( -2\alpha+(d-3)\beta-2 \frac{e^{(2\alpha - (d-1)\beta)\phi} \beta\, \rho^2}{g_{xx}^{d-1}} \right) \varphi  \ ,
\end{align}
and three non-independent first-order constraints  from the radial components of the Einstein equations:
\begin{align}\label{eq.app.scal7}
0 = & q\, H_{01}' + \omega \left( H_{11}' + (d-2) H_{ii}' \right) + q\, \partial_r \log \left[ \frac{g_{xx}}{g_{tt}} \right] H_{01} + \omega\, \partial_r \log \left[ \sqrt{ \frac{g_{xx}}{g_{tt}} } \right] H_{11}  \nonumber \\
& + \omega\, \partial_r \log \left[ \left( \frac{g_{xx}}{g_{tt}} \right)^\frac{d-2}{2} \right] H_{ii} + \omega\, \phi'\, \varphi \ ,
\end{align}
and
\begin{equation}\label{eq.app.scal7b}
0 =  q\, H_{00}' - \omega \frac{g_{xx}}{g_{tt}} H_{01}' + q (d-2) H_{ii}' + q\, \partial_r \log \left[ \sqrt{ \frac{ g_{tt} }{ g_{xx} } } \right] H_{00}  -T_b\, \lambda\, \rho \sqrt{ \frac{ g_{rr} }{ g_{tt}g_{xx}^{d-1} } } \left( \omega\, \alpha_1 + q\, \alpha_0 \right) +q\, \phi'\, \varphi \ ,
\end{equation}
and
\begin{align}\label{eq.app.scal8}
0 & = \partial_r \log \left[ g_{xx}^{d-1} \right] H_{00}' + \partial_r \log \left[ g_{tt} g_{xx}^{d-2} \right] \left(H_{11}' + (d-2) H_{ii'} \right)  + 4\,q\,\omega \frac{g_{rr}}{g_{tt}} H_{01} \nonumber \\
&  - 2 q^2 \frac{g_{rr}}{g_{xx}} H_{00}  + 2 \omega^2 \frac{g_{rr}}{g_{tt}} H_{11}  + 2 T_b\, \lambda\, \rho\, g_{rr}\frac{Z_1^{-2} Z_2^{-(d-1)} \rho^2 + g_{xx}^{d-1}}{\sqrt{g_{rr}g_{tt}} \, g_{xx}^{\frac{3}{2}(d-1)} } \alpha_0'  - 2 \phi' \varphi' \nonumber \\
& + 2 (d-2)g_{rr} \left(  \frac{\omega^2}{g_{tt}} - \frac{q^2}{g_{xx}} \right) H_{ii} - T_b \, \rho^2 g_{rr} \frac{\sqrt{ \rho^2 + Z_1^2 Z_2^{d-1} g_{xx}^{d-1} }}{Z_1^{2} Z_2^{d-1}  g_{xx}^{\frac{3}{2}(d-1)}} \, H_{00} \nonumber \\
& + g_{rr} \Bigg( 2\,  \partial_\phi V - T_b\,  Z_2^{d-2}\frac{\sqrt{ \rho^2 +Z_1^2 Z_2^{d-1} g_{xx}^{d-1} } }{g_{xx}^{\frac{3}{2}(d-1)}}  \Big( 2 e^{2\alpha\phi}\beta \, \rho^2 + e^{(d-1)\beta\phi} (2\alpha - (d+1)\beta)g_{xx}^{d-1}  \Big)  \Bigg) \varphi \ .
\end{align}

\end{document}